\documentclass[usenatbib]{mn2e}
\usepackage{times}
\usepackage{amssymb}
\usepackage{amsmath}
\usepackage{graphicx, natbib}
\usepackage{rotating}
\usepackage{lscape}
\usepackage{multirow}
\usepackage{ulem}
 
\input{mmp.abrev}
 
\title[]{Precision cosmology with a wide area XMM cluster survey}
\author[M. Pierre et al ]{M. Pierre$^{1}$\thanks{E-mail:mpierre@cea.fr}, F. Pacaud$^{2}$, J.B. Juin$^{3}$, J.B. Melin$^{4}$, P. Valageas$^{5}$,
N. Clerc$^{1}$, P.S. Corasaniti$^{6}$\\
$^{1}$DSM/Irfu/SAp, CEA/Saclay, F-91191 Gif-sur-Yvette Cedex, France\\
$^{2}$Argelander Institut f\"ur Astronomie, Universit\"at Bonn, Germany \\
$^{3}$Departamento de Astronom\'ia y Astrof\'isica, Pontificia Universidad Cat\'olica de Chile, Casilla 306, Santiago 22, Chile\\
$^{4}$DSM/Irfu/SPP, CEA/Saclay, F-91191 Gif-sur-Yvette Cedex, France\\
$^{5}$DSM/IPhT, CEA/Saclay, F-91191 Gif-sur-Yvette Cedex, France\\
$^{6}$CNRS, Laboratoire Univers et Th\'eories (LUTh), UMR 8102 CNRS, Observatoire de Paris,
Universit\'e Paris Diderot, 5 Place Jules Janssen, 92190 Meudon, France}
\begin{document}
\normalem
\newcommand{\dd}{deg$^{2}$}

\date{}
\maketitle

\label{firstpage}

\begin{abstract}

We explore the cosmological constraints expected from wide area XMM-type cluster surveys covering 50-200 \dd, under  realistic observing conditions.
We perform a Fisher matrix analysis based on cluster number counts in combination with
estimates of the $2$-point cluster correlation function. The effect of the survey design is implemented 
through an observationally well tested cluster selection function.
Special attention is given to the modeling of the shot noise and sample
variance, which we estimate by applying our selection function to
numerically simulated surveys. We then infer the constraints  on the equation of state of the dark energy considering various survey configurations. We quantitatively investigate  the respective impact of the cluster mass measurements, of the correlation function and of the $1<z<2$ cluster population.
We show that, with some 20 Ms XMM observing time, it is possible to constrain the dark energy parameters at a level which
is comparable to that expected from the next generation of cosmic probes. Such a survey has also the power to provide unique
insights into the physics of high redshift clusters and AGN properties.
\end{abstract}


\section{Introduction}
The statistical properties of galaxy clusters provide independent cosmological information,
complementary to that inferred from other observations such as measurements of the Cosmic Microwave 
Background (CMB), Supernova Type Ia (SN Ia), 
Baryon Acoustic Oscillations (BAO) and weak lensing (WL) data. 
Clusters are the largest virialized objects (dark matter halos) in the Universe, 
with mass scales corresponding to overdensities that enter the non-linear 
phase of gravitational collapse between redshifts $0<z<3$. 
Consequently, their abundance and spatial distribution can potentially probe both the 
cosmic expansion history as well as the growth of cosmic structures.  
Theoretical considerations such as the prediction of 
the halo mass function based on semi-analytical approaches \citep{PressSchec74,Bond91} 
and N-body simulations \citep[see e.g.][]{ShethTor99} have suggested 
that cluster statistics is particularly sensitive to the normalization of the 
matter power spectrum $\sigma_8$ (the root-mean-square of linear fluctuations within 
a sphere of $8 h^{-1}$ Mpc radius) and the total cosmic matter
density $\Omega_{\rm m}$. These observational aspects 
have given a strong incentive to the use of clusters as cosmic probes.

Over the past decades cluster observations have greatly evolved. After the pioneering studies 
of the Einstein Medium Sensitivity Survey \citep{Gioia90}, the Rosat All-sky survey (RASS) 
and deep ROSAT pointed observations have provided an invaluable reservoir 
of clusters out to redshift $\sim1$. These measurements enabled the first determinations of 
$\sigma_{8}$ and $\Omega_{\rm m}$ based on cluster number counts alone 
\citep[see][]{Evrard89,Ouk92,White93,Viana96,Eke98,Henry97,Henry00,Borgani01,Vikh03,allen03} 
and in combination with measurements of the local correlation function from RASS \citep{schuecker03}. Similarly
the Sloan Digital Sky Survey cluster catalogue offered the 
first determination using an optical dataset \citep[e.g.][]{Bahcall03}.
Quite remarkably these measurements have always consistently pointed out
to a low matter density universe, in agreement with results from galaxy survey data \citep{Percival01,Tegmark04} and 
CMB observations \citep{DeBernardis00,Spergel03}. 
With the launch of XMM and Chandra a decade ago, a new era has begun: deep pointed observations 
of large cluster samples, mainly extracted from the ROSAT catalogues, have provided detailed insights 
into the baryonic physics of clusters and their morphology. This has resulted in a tremendous burst in the modeling 
of the cluster properties as well as in the determination of their mass. These advancements have led to
improved constraints on $\sigma_{8}$ and $\Omega_{\rm m}$, as obtained for example 
using the temperature function of local bright clusters \citep{henry09}. From the point of view of large area surveys, the XMM-LSS survey \citep{Pierre04} covering some 11 \dd\ performed pioneering cluster detection work, assembling a complete sample of  XMM clusters at a sensitivity of $\sim 10^{-14}$ \flux\ in the [0.5-2] keV band. Moreover, it provided detailed insights about the impact of selection effects on cluster evolutionary studies \citep{pacaud07}.

The discovery of dark energy has generated a revived interest in the use of cluster statistics as 
an alternative test for probing the nature of this exotic component. Dark energy can directly 
affect the cluster number counts by modifying the growth rate of structures as well as 
the size of the cosmological volume probed at a given redshift 
\citep{WangStein98,Haiman01,HutTurn01}. Several works have attempted to measure the dark energy equation of state
using cluster data in combination with other probes \citep{Henry04,Mantz08,Vikhlinin09,allen08,Rozo09}. 
However statistical and systematic uncertainties, as well as the presence of degeneracy between cosmological
parameters, remain the major limitations to accurately test dark energy with current data.
 
From an observational point of view, the main quantities that are useful to constrain cosmology are: 
the redshift evolution of the cluster number counts ($dn/dz$) or ideally the evolution of the 
cluster mass function ($dn/dMdz$), the spatial distribution of clusters 
(e.g. the two-point correlation function, $\xi$), the cluster temperature function, 
the gas mass fraction in clusters as well as various scaling laws describing 
the evolution of cluster structural properties. 
There are two key practical issues that such studies have to face: firstly the ability to assemble 
well characterized cluster samples, and secondly the need for well understood mass-observable relations, 
since for a given cosmology the cluster mass is the only independent variable entering the theory. 
Mass estimates can be inferred from a variety of methods: 
optical richness, galaxy velocity dispersion, X-ray luminosity or temperature,
S-Z decrement, weak lensing signal or from more elaborated proxies such as $T_{\rm X} \times M_{\rm gas}$ 
described in \citep{kravtsov06}; if X-ray temperature and gas density profiles are available, masses can 
be calculated under the hypothesis of hydrostatic equilibrium.

Depending on the number of cosmological parameters that one aims at constraining and the required accuracy, 
the minimum size of useful cluster samples ranges from 50-100 objects for constraining $\sigma_{8}$
and $\Omega_{\rm m}$ only, to several hundreds or even several thousands (if little information 
is available on masses) for constraining the dark energy parameters.

Since clusters constrain regions of the cosmological parameter space which are complementary to that 
probed by other tests such as SN Ia, CMB, BAO and WL data \citep[see e.g.][]{HutTurn01},
considerable efforts have been devoted, both theoretically and observationally,
to characterize the use of clusters in the near future. 
Forecasts of the dark energy parameter uncertainties
from future optical, X-ray and S-Z surveys have been the subject of several analyses
\citep{Weller02,HuKrav03,Mohr03,Mohr04,Wang04,Wu08}. 
These studies, generally focusing on surveys covering a few 1 000 \dd, have shown that precision cosmology in the context of cluster surveys is certainly possible in the near future. Subsequently, there has been a growing interest in evaluating  the impact of systematic uncertainties of such cluster surveys. For instance, one can mention  the sensitivity  of the dark energy constraints to halo modeling uncertainties \citep{Cunha09} or to the mass accuracy of given cluster sub-samples; the latter is of special relevance when designing the follow-up observations to increase the cluster mass accuracy: given that telescope time is limited, it is necessary to optimise the targeting of specific mass and redshift ranges  \citep{Wu10}. 

While these prospective dark energy  studies pertain to upcoming or future instrumentation, we examine here the potential of XMM, whose characteristics and capabilities are now very well established.
In fact, with its outstanding collecting area 
($\sim2000 ~{\rm cm^{2}}$ on axis at 1 keV), wide spectral range ([0.1-10] keV), good spatial 
($\sim6$ arcsec on axis) and spectral (5-10\% at 1 keV) resolution, 
XMM appears to be the best suited, currently available, X-ray observatory 
to undertake a large cluster survey. As an example, with 10 ks exposures, 
XMM reaches a sensitivity which is about 1000 times greater than RASS,  
i.e. $5\times 10^{-15} {\rm erg ~cm^{-2} ~s^{-1}}$ in [0.5-2] keV for point sources. Basically, XMM has the power 
to unambiguously resolve any cluster\footnote{A core radius of 150 kpc corresponds to an apparent diameter of  35 arcsec at $z=2$, to be compared to the XMM on-axis PSF of 6 arcsec} provided 
that at least some $100$ photons are collected.

In this paper, we forecast the dark energy parameter errors for an XMM cluster survey with an area
of the order of 100 \dd. 
Using results from accurate survey simulations and precise model predictions, we estimate the
dark energy parameter errors for different survey configurations. We find that the expected
parameter constraints are not only complementary to those of other cosmological probes, but
competitive with respect to forecasted errors for the next generation of dark energy dedicated experiments.

Compared with other cluster surveys, X-ray observations have an indisputable advantage,
since cluster X-ray properties can be predicted {\em ab initio} for a given cosmological model, 
with {\em observational input} (e.g., mass-observable relations) being easily implementable. 
In contrast, ground-based large optical cluster surveys \citep[e.g., SDSS Max BCG catalogue,][]{Koester07}, though
 may appear much more attractive because of their lower cost, 
still require {\em ad hoc} prescriptions to evaluate the cluster selection function
with cosmological numerical simulations. Such procedures usually rely on the optical 
richness as defined by the galaxy distribution. 
We want to stress that computing a cluster survey selection function in the era of precision 
cosmology requires a self-consistent modeling of the selection
function itself. We will show here that this plays a critical role in the interpretation of the cluster number counts. 
It is also worth mentioning that, after 40 years of experience, X-ray cluster surveys  
are still much ahead of S-Z surveys both in terms of detection rates and for the
evaluation of the selection function. In the following we shall refer to the discussed survey as the XXL survey.\\

The paper is organized as follows. In Section~\ref{ClustObs} we
introduce the basic equations for the cluster survey
observables, namely the cluster number counts and the $2$-point correlation
function. In Section~\ref{xxlsurvey} we describe the survey
configurations and selection functions, while in
Section~\ref{surveyerr} using numerical simulations we estimate the
expected experimental survey uncertainties. 
In Section~\ref{Fisher-analysis} we describe the Fisher matrix
calculation performed to infer the expected cosmological parameter constraints, 
and discuss the results in
Section~\ref{constraints}. Finally we present our summary and conclusions in Section~\ref{conclusion}.
Throughout the paper,  we consider the $\Lambda$CDM cosmology with the parameters determined by WMAP-5 \citep{Dunkley09} as our fiducial cosmological model. 
 
\section{Cluster Survey Observables}\label{ClustObs}
The number of clusters as function of redshift is given by
\begin{equation}
\frac{dn}{dz}=\Delta\Omega \frac{d^2V}{d\Omega dz}(z)\int_0^{\infty}F_{\rm s}(M,z)\frac{dn(M,z)}{d\log{M}}d\log{M},\label{counts}
\end{equation}
where $\Delta\Omega$ is the survey solid angle, $d^2V/d\Omega dz$ is the cosmological volume factor, 
$F_{\rm s}(M,z)$ is the redshift dependent survey selection function and $dn/d\log{M}$ is the comoving density of halos
of mass $M$. 

The volume factor in a Friedmann-Robertson-Walker space-time reads as 
\begin{equation}
\frac{d^2V}{d{\Omega}dz} = \frac{c}{H_0} \frac{(1+z)^2 d_{\rm a}^2(z)}{E(z)} ,
\end{equation}
with $c$ the speed of light, $H_0$ the Hubble constant today, and 
\begin{equation}
E(z)\!=\!\sqrt {\Omega_{\rm m}(1\!+\!z)^3+\Omega_{\rm r}(1\!+\!z)^4+\Omega_{\rm DE} I_{\rm DE}(z)
+\Omega_k (1\!+\!z)^2},
\end{equation}
where $\Omega_{\rm m},\Omega_{\rm r},\Omega_{\rm DE},\Omega_k$ are the present matter,
radiation, dark energy, and curvature densities,
in units of the critical density respectively, and $d_{\rm a}(z)$ is the angular diameter distance. 
The function $I_{\rm DE}(z)$ depends on the model of dark energy. 
We consider three scenarios: 1) cosmological constant $\Lambda$, with $I_{\Lambda}(z)=1$; 
2) dark energy fluid characterized by a constant equation of state $w$ for which $I_{\rm DE}(z)=(1+z)^{3(1+w)}$;
3) time evolving dark energy equation of state parametrized in the form 
$w_{\rm DE}(z)=w_0+w_az/(1+z)$ \citep{Polarski,Linder} for which $I_{\rm DE}(z)=(1+z)^{3(1+w_0+w_a)}\exp{[-3 w_a z/(1+z)]}$.

Cluster DE studies make various assumptions as to the selection function. It can be defined by a simple mass limit, depending or not on redshift and  cosmology; the limit is supposed to be step-like or  to allow for a  possible dispersion and for some smooth function across the threshold \citep[e.g.][]{Lima05, Hu06, Albrecht06, Basilakos2010}. For an X-ray survey,
$M_{\rm lim}(z)$ is determined using a mass-observable relation, for instance the mass-luminosity relation.
Because of this, the limiting mass depends on empirically determined parameters
(parametrizing the physics of the hot gas in clusters), and the luminosity distance, i.e.
on the underlying cosmological model. This is an important point, since it implies that to properly
infer cosmological constraints from cluster data one must take into
account the cosmological dependence 
of the mass threshold $M_{\rm lim}(z)$. Another relevant aspect concerns the fact that survey
design, instrumental characteristics and imaging technique analysis, cause the cluster 
selection function not to behave as a step function. This can affect 
the predicted number counts and 2-point correlation statistics for a given cosmological model. 
The effect of a non step-like function has been studied in the literature for
S-Z surveys \citep{Holder00,LinMo03} and X-ray surveys \citep{Sahlen09}. Here we use realistic selection functions derived 
from accurate simulations of the XMM-LSS survey that will be discussed in Section~\ref{xxlsurvey}.

The comoving density of halos of mass $M$ at redshift $z$ reads as 
\begin{equation}
\frac{dn(M,z)}{d\log{M}}=-\frac{\bar{\rho}_{\rm m}}{M}\frac{d\log{\sigma}}{d\log{M}}f(\sigma,z),
\end{equation}
with $\bar{\rho}_{\rm m}$ the present mean matter density, $\sigma(M,z)$ the root-mean-square fluctuation 
of the linear density contrast smoothed on a scale $R=(3M/4\pi\bar{\rho}_{\rm m})^{1/3}$, 
and $f(\sigma,z)$ the multiplicity function. Here we adopt for $f(\sigma,z)$ the modeling proposed by \citet{Tinker08}.
Our working assumptions are detailed in Appendix \ref{appendixA}.\\
The variance of the linear density contrast smoothed on scale $R$ at redshift $z$ is given by
\begin{equation}
\sigma^2(R,z)=A^2\int\frac{dk}{2\pi^2} k^{n_s+2}T^2(k,z)W^2(kR),
\end{equation}
where $A$ is a normalization constant fixed so that today $\sigma(R=8 h^{-1}{\rm Mpc})=\sigma_8$, 
$n_s$ is the scalar spectral index, $T(k,z)$ is the linear matter transfer function and $W(kR)$
is the Fourier transform of the real space top-hat window function. We compute the matter transfer function using the fitting formula provided by \cite{EisenHu98}, which includes the wave pattern imprinted by the baryon acoustic 
oscillations.\\
On large scales the $2$-point spatial correlation function for a cluster survey covering the redshift range 
$[z_{\rm min},z_{\rm max}]$ is given by
\begin{equation}
\xi(R)=\frac{\int^{z_{\rm max}}_{z_{\rm min}} \frac{d^2V}{d\Omega dz} n^2(z)\xi(R,z)dz}{\int^{z_{\rm max}}_{z_{\rm min}}\frac{d^2V}{d\Omega dz}n^2(z)dz},\label{csi}
\end{equation}
where 
\begin{equation}
n(z)=\int_0^{\infty}F_{\rm s}(M,z)\frac{dn(M,z)}{d\log{M}}d\log{M},
\end{equation}
and $\xi(R,z)=b_{\rm eff}^2(z)\xi_{\rm lin}(R,z)$, with $\xi_{\rm lin}(R,z)$ the Fourier transform of the
linear matter power spectrum at redshift $z$.
The evolution of the linear bias averaged over all halos reads as \citep{Matarrese97}
\begin{equation}
b_{\rm eff}(z)=\frac{1}{n(z)}\int_0^{\infty}F_{\rm s}(M,z)b(M,z)\frac{dn(M,z)}{d\log{M}}d\log{M},
\end{equation}
where $b(M,z)$ is the linear bias relating dark matter halos of mass $M$ to the mass density fluctuation.
We assume the bias model introduced in \cite{Tinker10},
\begin{equation}
b(M,z)=1-\frac{1+A_b}{1+\sigma^{a_b}}+0.183\left(\frac{\delta_c}{\sigma}\right)^{1.5} + B_b\left(\frac{\delta_c}{\sigma}\right)^{2.4},
\end{equation}
with $\delta_c=1.686$ the critical linear overdensity given by the spherical collapse model,
and the fitting parameters given by
\begin{align}
A_b=0.24y \exp{[-(4/y)^4]}, & \hbox{\hspace{0.5cm}}\\
a_b=0.44(y-2), &  \\
B_b=0.019 + 0.107y + 0.19\exp{[-(4/y)^4]}, & 
\end{align}
where $y= \log_{10}(\Delta_{\rm m})$ with $\Delta_{\rm m}$ the nonlinear overdensity
threshold. Our fixed value of $\delta_c$ is only exact for an 
Einstein-de Sitter universe - although it hardly varies with the cosmology. 
Nevertheless, we preferred to follow the convention of \cite{Tinker10} and fix it.

\section[]{XXL Survey Characteristics}\label{xxlsurvey}

Cluster surveys are, similarly to galaxy surveys, defined by a number of parameters such as sky coverage 
and geometry, depth, selection function, and redshift accuracy. On the other hand, compared to galaxies, 
clusters are rare objects, a characteristic that has a significant impact on the determination of the correlation function. 
Moreover, as already mentioned in the introduction, cluster mass accuracy plays an important role in 
the determination of the cosmological parameters. In this section, we
present the generic characteristics of the XXL survey,
while a quantitative examination of the various sources of uncertainty will be presented 
in Section~\ref{surveyerr}.

\subsection{Two survey designs}

In this case study, we examine the merits of two possible XMM survey concepts:  
{\sl Survey-{A}} covers a total sky area of 50 \dd\ with 40 ks XMM pointings, this configuration is assumed 
to allow mass measurements at the 10-50\% level for
the selected cluster samples;  
{\sl Survey-B }  covers 200 \dd\ with 10 ks XMM pointings and provides a cluster mass accuracy of 50-80\%.
Possible survey configurations, resulting from various splitting in sub-regions, are summarized 
in Table \ref{simul} and discussed in Section~\ref{estim-noise}.

To give an order of magnitude of the observing time necessary to perform these surveys, one can imagine 
mosaics consisting of XMM observations whose center are separated by 20 arcmin in RA and Dec, so that 9 observations 
are necessary to cover 1\dd. Consequently, both surveys A and B correspond approximately to $\sim$ 18 Ms net observing time, 
i.e. some 180 XMM (2-day) revolutions, allowing for 10 observations per revolution with the mosaic mode.

\subsection{Modeling the cluster population as seen by XMM}

Before detailing the survey selection function, we need to specify how the two basic observable 
quantities, the X-ray count rate in a given band and the apparent size of the cluster sources, relate to the cluster mass as a function of redshift. \\

In the following, we assume the usual [0.5-2] keV range as the working detection band, since it presents 
the optimal S/N, given the cluster spectra, the background spectrum and the XMM spectral response \citep{Scharf02}.
Furthermore we assume the observed cluster scaling laws between luminosity ($L$), temperature ($T$) and the 
mass within a radius containing an overdensity of 200 times the critical density ($M_{200c}$) 
as determined in the local universe \citep{ArnaudEvr99,Arnaud05},  and use the self-similar 
prescription for their evolution. To account for the scatter observed in cluster properties, we
encapsulate the dispersion of the $M-T$ and $L-T$ relations in the $M-L$
relation, for simplicity. Following the analysis by \cite{stanek06}, 
who measured $\sigma_{\ln{M}|L}=0.37$, we use $\sigma_{\ln{L}|M}\sim0.37\times1.59\sim0.6$, 
where $1.59$ is the slope of their $M-L$ relation. To assign the X-ray luminosity we assume a log-normal distribution. These prescriptions allow us to compute the flux, and finally the count rate as function of the cluster mass and redshift. The impact of these hypotheses will be discussed in Sec. \ref{constraints} and \ref{conclusion}. 

Fluxes are estimated using the APEC thermal plasma model, assuming a fixed hydrogen
column density of $2.6\times 10^{20}$~cm$^{-2}$ and setting the heavy element abundance
to 0.3 solar. 
Fluxes are subsequently folded with the   telescope and detector response (EPIC response matrices) assuming the THIN optical blocking filter. This allows us to  predict the 
observed count-rates. We further assume a $\beta$-profile for the gas distribution, with $\beta$=2/3 and a 
constant physical core radius of $180$~kpc, unless otherwise specified. This finally yields the spatial distribution 
of the cluster counts on the  detectors.

\subsection{The cluster selection function}

\label{cluster-select}
We now turn to the description of the selection function. The ability to select clusters upon well-defined X-ray 
criteria is a key issue: as shown in Section~\ref{ClustObs}, the selection function directly 
enters into the modeling of the cluster number counts and spatial correlation function.\\

\begin{figure}
\includegraphics[width=8cm]{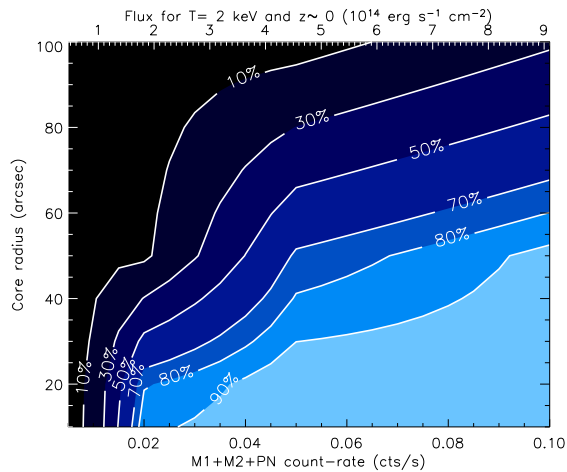}
  \caption{The C1 cluster selection function derived from extensive
    simulations: the probability of cluster detection is expressed in
    the count rate ($\sim$ flux ) - core radius plane. A $\beta$-model with $\beta=2/3$ is assumed.}
\label{c1sel}  
\end{figure}

\begin{figure}
\includegraphics[width=8cm]{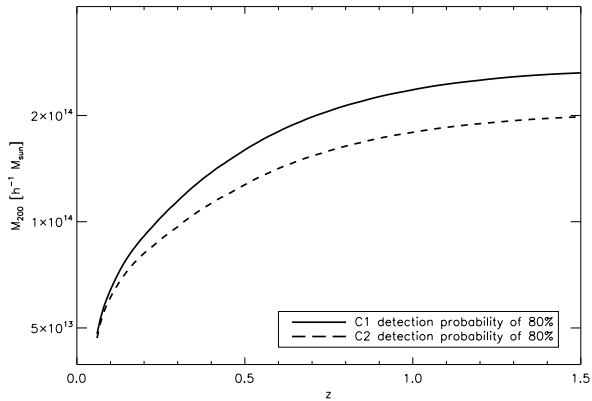}
 \caption{The limiting detectable cluster mass as a function of
   redshift. A detection probability of 80\% is assumed. Masses 
are expressed in terms of $M_{200c}$, the mass within a radius
   containing an overdensity 200 times the {\em critical density}.}
 \label{mlim}
\end{figure}

\begin{figure}
\includegraphics[width=8cm]{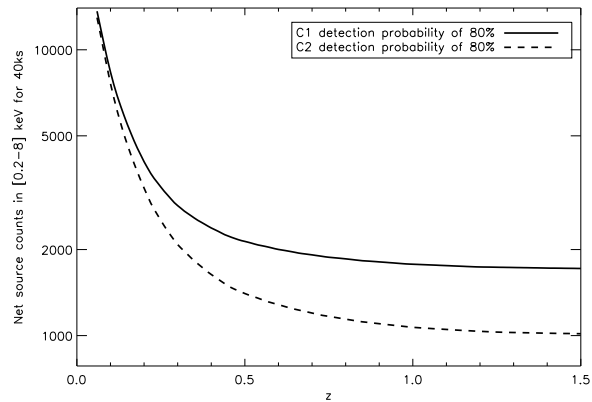}
 \caption{Number counts (2 MOS + pn) collected in 40 ks ({\sl  Survey-A} configuration) from a C1 and
   C2 cluster in the [0.2-8] keV energy range, as a function
   redshift. A detection probability of 80\% is assumed, thus
   corresponding to the $M_{\rm lim}(z)$ of Fig.~\ref{mlim}. The EPIC
   sensitivity has been averaged over the inner $r=10~{\rm arcmin}$
   (mean vignetting of 0.69). Assuming that half of the collected
   photons are used for the spectral analysis, our selection ensures
   that at least 500 counts are available for temperature determination with 40 ks XMM exposures.}
 \label{clim}
\end{figure}

In this prospective study, we adopt the C1/C2 selection functions
specifically determined  for the  XMM-LSS survey. 
These have been extensively tested on the basis of XMM image simulations \citep{pacaud06} and 
applied  to the XMM-LSS sample \citep{pacaud07}. The selection basically operates in the 
{\sl  [extent, extent likelihood]} X-ray pipeline parameter space\footnote{Because of the limited number of source photons, the pipeline operates in Cash statistics and returns, for each source parameter,  the likelihood of the measurement},
where  {\sl extent}  is taken to be the core radius of the  $\beta$-model. The procedure allows us to assemble
samples of extended X-ray sources that have a well-defined degree of contamination by miss-classified point-source; 
these can be easily discarded a posteriori by examining the X-ray/optical overlays.  We define two samples, 
C1 and C2, for which the contamination is  $\sim$ 0 and $\sim$ 50\% respectively \citep{pierre06}.  This procedure, which operates in a two-dimensional parameter space, enables 
the construction of uncontaminated cluster samples significantly larger than those obtained by a simple flux limit. 
The selection criteria are subsequently converted into the probability of detecting a source characterized 
by a given core radius and flux. The C1 selection probability function is displayed in Fig. \ref{c1sel}.  
Using the cluster model described in the previous paragraph, we derive the limiting cluster mass 
detectable as a function of redshift for C1 and C2 respectively. 
Since the current C1/C2 selection criteria have been defined for 10 ks XMM exposures, the resulting selection 
corresponds to clusters having $M_{200c} > 2 \times 10^{14} ~ M_\odot$, thus relatively massive objects as can
be seen in Fig.~\ref{mlim}. 
Moreover we note that $M_{\rm lim}(C1)\sim 1.5 \times M_{\rm lim}(C2)$ for $z>0.2$, with  the C2 selection yielding about 
twice as many clusters as the C1 selection. Notice that the C1 sample is always a sub-sample of the C2 selection.
The number of collected cluster counts at the detection limit is displayed in Fig.~\ref{clim}.\\

\begin{figure}
\includegraphics[width=8cm]{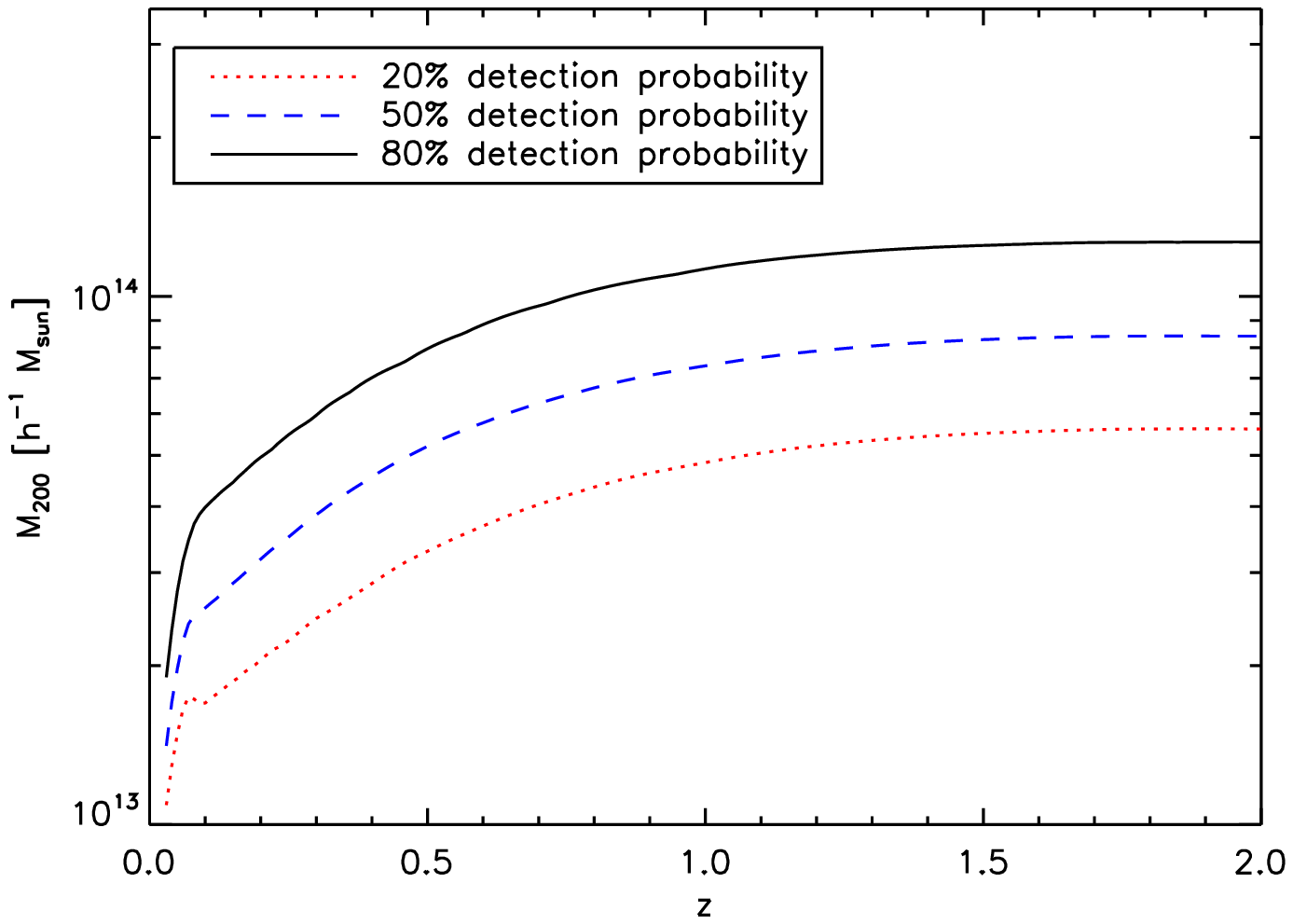}
 \caption{Same as Fig. \ref{mlim} for the C20 population}
 \label{mlimc20}
\end{figure}

\begin{figure}
\includegraphics[width=8cm]{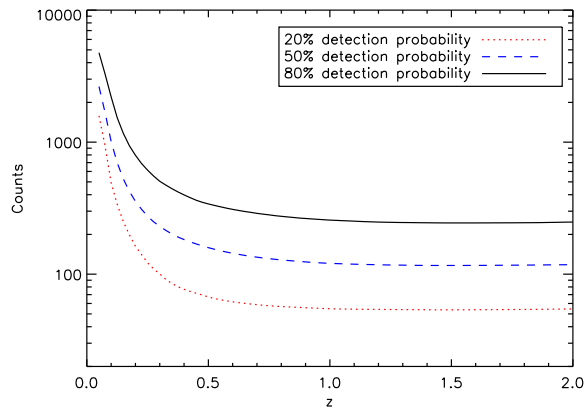}
 \caption{Same as Fig. \ref{clim} for the C20 population }
 \label{climc20}
\end{figure}

Practically, our cosmological analysis will be performed in two stages.
(i) In a first step, we  consider the same cluster selection functions independently of the survey configuration (A or B). This means that for configuration A,  the sample is defined from sub-exposures of 10 ks. The main goal of the total 40 ks integration time 
is to reach the X-ray spectral accuracy  enabling accurate mass measurements. Further, at the full depth of 40 ks,  {\sl Survey-{A}} enables the detection of deeper cluster samples. Consequently (ii) in a second step, we investigate the added cosmological value from clusters only detected in the  40 ks observations of  {\sl Survey-{A}}.  We thus define a C20 class, a scaled-down version of the C2 population detected in 10 ks.  Since the C2 selection function is well depicted by a 
detection probability as a function of S/N, we simply derived the C20 detection 
efficiency by extrapolating the results of Pacaud et al. (2006) to 40ks, scaling up the 
source S/N\footnote{ This method was already  applied in Pacaud et al. 
(2007) to account for the spatial variations of exposure time.}. The density inferred for this population is on the order of 30/\dd and comparable to that inventoried in the $\sim$ 50ks COSMOS field by \citet{Finoguenov07}. The characteristics of the C20 clusters are displayed in Figs. \ref{mlimc20} and \ref{climc20}.  The number densities of the C1, C2, C20 populations are given in Table \ref{C1C2C20density}. Furthermore, we define the following sub-classes:
 we refer to C2' for C2 clusters not detected as C1 and, similarly, to C20' for the C20 clusters not detected as C2.

\begin{table}
 \centering
  \caption{Properties of the cluster samples selected for the cosmological analysis}
  \begin{tabular}{@{}lccc@{}}
  \hline \hline
  Selection        &     Detected in configuration    & \multicolumn{2}{c}{Number density (deg$^{-2}$)}    \\
  ~ & ~ & $z<1$ & $z<2$ \\  \hline
   C1 & A ~~ B & 7.1 & 8.0 \\
    C2 & A ~~ B & 11.6 & 13.7 \\
 C20 & A  & 23.2 & 28.2 \\
 \hline \hline
\end{tabular}
\label{C1C2C20density}
\end{table}

\section[]{Estimating measurement uncertainties}\label{surveyerr}

We provide in this section a detailed account of the uncertainties pertaining to the measurements of clusters masses, cluster number counts and $2$-point correlation function, as expected from the XXL survey. These are the necessary ingredients for a 
realistic evaluation of the cosmological parameter errors  via a Fisher matrix analysis.
\subsection{Accuracy of the cluster mass and redshift measurements}

For this study, we do not rely on the, 
so far non-observationally validated, self-calibration techniques that allow for some universal redshift-dependent mass-observable relation  \citep{Mohr04}; we discuss the relevance of this 
option in Sec. \ref{conclusion}. We rather attribute to each cluster a mass accuracy as a function of its X-ray flux. The limiting collected counts for the three cluster populations, as indicated by Figs. \ref{clim} and \ref{climc20}, allow us to estimate the mass accuracy reachable for each selection.  We consider a pessimistic and an optimistic situation and further set a limit on the precision of the  observed luminosities. These working hypotheses are listed in Table \ref{masserr}.
It is not the purpose of the present article to discuss in detail how such 
mass accuracy will be obtained, but one can foresee a set of realistic observations leading to the desired precision. 
For instance, configuration B is similar to the well studied XMM-LSS design, i.e. a mosaic of 10 ks exposures, 
which allowed mass measurements to better than 50\% for the C1 population,  under the assumption of hydrostatic equilibrium  \citep{pacaud07}. Improved cluster mass accuracy will be attained with the addition 
of weak lensing and Sunyaev-Zel'dovich observations (e.g. \citet{mahdavi07}). The use of several X-ray mass proxies, such as the 
$Y_{\rm X} = T\times M_{\rm gas}$ parameter, can also greatly enhance the precision of the mass estimates \citep{Vikhlinin09}. 

\begin{table}
 \centering
  \caption{Adopted mass precision for each individual cluster as a function of XMM exposure time. The  numbers are the 1-$\sigma$ errors on $\ln(M)$. The  * indicates that this sub-population does not provide mass information for the Fisher analysis.  Last line gives the assumed precision on the luminosity measurements.
 }
  \begin{tabular}{@{}lcccc@{}}
  \hline \hline
  Selection         & \multicolumn{4}{c}{Adopted mass accuracy}    \\
  ~ &  \multicolumn{2}{c}{Optimistic view}   & \multicolumn{2}{c}{Pessimistic view}    \\ 
  ~ & 10 ks & 40 ks &10 ks &40 ks \\ \hline \hline
   C1 &  0.5 & 0.1& 0.8 & 0.5 \\
    C2' & 0.8 &0.5 &  * & 0.8 \\
 C20' & not detected & 0.8 & not detected & * \\ \hline
 $\sigma_{\ln L_{obs}}$ & \multicolumn{2}{c}{negligible} &  \multicolumn{2}{c}{0.2} \\  \hline \hline
\end{tabular}
\label{masserr}
\end{table}

Experience with the Canada France Hawaii Telescope Legacy Survey (CFHTLS\footnote{http://www.cfht.hawaii.edu/Science/CFHLS/}) showed that cluster photometric redshifts can be obtained for the C1 and most of the C2 clusters  at an accuracy of $\sim 0.01-0.02$ from a 5-band survey in the optical \citep{Mazure2007}. Further, with 
the up-coming generation of wide-field spectroscopy instruments (e.g. refurbished VIMOS and forthcoming KMOS at 
the ESO Very Large Telescope) gathering redshifts of clusters with a density of $\sim10-50/{\rm deg^2}$ over an area of 100 \dd\ will be easily achievable within the next decade.

\subsection{Statistical significance of $dn/dz$ and $\xi$}
\label{estim-noise}
Evaluating the impact of the survey size on the statistical significance of $dn/dz$ and $\xi$ from cluster
surveys deserve special attention. Because clusters are rare objects,  the relative effects of shot noise, 
sample variance and edge effects as functions of the survey depth and geometry are quite different from that of 
galaxy or weak lensing surveys. More precisely, considering splitting the survey in several sub-regions (a strategy favored by practical observing considerations), we need to estimate the trade-off  between  averaging the sample variance  and the loss of S/N in the 2-pt correlation function at large distances.
In principle, it is possible to analytically calculate the sample variance and the shot noise  for $dn/dz$ and $\xi$ as a function of cosmology for a given flux limited or volume limited survey \citep[e.g.][]{HuKrav03}. Having here a well defined selection function $M_{\rm lim}(z)$, we perform an ``in situ'' and global estimate using numerically simulated cluster samples. The corresponding calculations are detailed in Appendix \ref{appendixB} \\

\section[]{Fisher matrix analysis}\label{Fisher-analysis}
We perform a Fisher matrix analysis to quantitatively estimate the cosmological information 
that can be extracted from the two XMM-survey configurations (A and B). 

\subsection{Method}
Here we briefly sketch the basic principle of the Fisher matrix approach, 
interested readers may find more exhaustive discussions on its cosmological applications
in \citep{Tegmark97,Eisenstein99}.

Let us consider a set of measurements $D_i=\{D_1,...,D_N\}$ (for simplicity let us assume them to be uncorrelated), from which
we want to derive constraints on a set of parameters $\theta_\mu=\{\theta_1,...,\theta_M\}$
in a given model $\mathcal{M}$. We firstly evaluate the likelihood function, $L(D_i|\theta_\mu,\mathcal{M})$, 
and assuming a prior probability distribution for the model parameters, $P(\theta_\mu|\mathcal{M})$, 
we construct using Bayes' theorem the posterior probability, i.e. the probability of the parameters given 
the observed data, $P(\theta_\mu|D_i,\mathcal{M})\propto L(D_i|\theta_\mu,\mathcal{M}) P(\theta_\mu|\mathcal{M})$. 
The posterior contains all statistical information from which 
we derive the ``confidence'' intervals on the parameters
$\theta_\mu$. Now, let us indicate with ${\rm O}_i(\theta_\mu)$ the
model prediction of the observable to be confronted with the data $D_i$, 
and let $\sigma_i$ be the experimental uncertainties. Assuming Gaussian distributed errors,
we can write up to an additive constant the log-likelihood as
\begin{equation}
\ln{L} = -\frac{\chi^2}{2}=-\frac{1}{2}\sum_{i=1}^N\frac{[D_i-{\rm O}_i(\theta_\mu)]^2}{\sigma_i^2}.\label{loglike}
\end{equation}
If $\hat{\theta}_\mu$ are the model parameter values that maximize the likelihood, then
we can expand Eq.~(\ref{loglike}) to second order in $\delta\theta_\mu=\theta_\mu-\hat{\theta}_\mu$ and obtain
\begin{equation}
\mathcal{L} \equiv - \ln{\left(\frac{L}{L_{\rm max}}\right)}= \frac{1}{4}\sum_{\mu,\nu=1}^M\left.\frac{\partial^2\chi^2}{\partial{\theta_\mu}\partial{\theta_\nu}}\right|_{\hat{\theta}} \delta\theta_{\mu}\delta\theta_\nu .
\end{equation}
This leads to the Fisher matrix $F_{\mu\nu}$ given by\footnote{Although we have assumed a Gaussian likelihood to derive this expression, 
it is worth noting that the Fisher matrix has exactly the same shape for Poisson statistics.}
\begin{equation}
F_{\mu\nu}\equiv\langle\frac{\partial^2\mathcal{L}}{\partial{\theta_\mu}\partial{\theta_\nu}}\rangle=\sum_{i=1}^N\frac{1}{\sigma_i^2}\left.\frac{\partial{\rm O}_i}{\partial\theta_{\mu}}\frac{\partial{\rm O}_i}{\partial\theta_{\nu}}\right|_{\hat{\theta}} .
\label{fisher}
\end{equation}
The parameter uncertainties as well as their mutual correlations are encoded in the covariance matrix, 
$C_{\mu\nu}=F_{\mu\nu}^{-1}$, where the $1\sigma$ model parameter errors  
are simply the square-root of the diagonal elements, $\sigma_{\theta_\mu}=\sqrt{C_{\mu\mu}}$.
These are the marginalized errors, in the sense that if we consider a
specific parameter, e.g. $\theta_1$, then the 
uncertainty $\sigma_{\theta_1}$ obtained by inverting the full Fisher
matrix is equivalent to that 
obtained by integrating the likelihood function over the $M-1$ parameters, thus
accounting for all possible parameter correlations. External priors on a given parameter can be easily implemented, e.g. suppose we want to
include a $\sigma_{\theta_{3}}=0.01$ prior on the parameter $\theta_3$, in such a case it is sufficient to add to Eq.~(\ref{fisher}) a matrix $P_{\mu\nu}$
whose only non-vanishing element is $P_{33}=1/\sigma_{\theta_3}^2$. Similarly information from other datasets can be easily implemented by
adding the corresponding Fisher matrices.

Using Eq.~(\ref{fisher}) greatly simplifies 
the estimation of the cosmological parameter uncertainties for a given experiment. 
Then forecasting parameter errors reduces to knowing the expected experimental/observational uncertainties ($\sigma_i$),
assuming a fiducial cosmology ($\hat{\theta}_\mu$) and computing the
Fisher matrix by evaluating the derivative 
of the observable at the fiducial parameter
values ($\partial{\rm O}/\partial\theta_\mu|_{\hat{\theta}_\mu}$). The inferred errors will necessarily depend on the 
fiducial cosmology assumed; this is the case even if one runs a full numerical likelihood
analysis over a set of randomly generated data. Henceforth the results of this type of analysis should not be used
for estimating the performance of experiments in distinguishing between different models. We refer the reader to \citep{Pia06} for
a discussion on the limitation of this approach in model selection problems and the solution in the context of Bayesian statistics.

We evaluate the derivatives of the observable with the respect to the model parameters using 
the five-point stencil approximation:
\begin{align}
\notag\frac{\partial{\rm O}}{\partial\theta_\mu} \approx & \ \ \frac{2}{3}\,\frac{O(\hat{\theta}_\mu+\delta\theta_\mu)-O(\hat{\theta}_\mu-\delta\theta_\mu)}{\delta\theta_\mu}\\
&  \ \ \ \ \ \ \ +\frac{{\rm O}(\hat{\theta}_\mu-2\delta\theta_\mu)-O(\hat{\theta}_\mu+2\delta\theta_\mu)}{12\delta\theta_\mu}\label{deriv}
\end{align}
with steps $\delta\theta_\mu$ of order $5\%$ on the fiducial parameter value.

Our survey observables consist of the cluster number counts $dn/dz$ given by Eq.~(\ref{counts}) in redshift bins of size 
$\Delta{z}=0.1$ and the two-point spatial correlation 
function $\xi(R)$ given by Eq.~(\ref{csi}). For the cluster counts we consider detections in $10$  or $20$ equally spaced redshift bins
in the range $0<z<1$ or $0<z<2$, while for the correlation function we consider the
$10<R~(h^{-1}{\rm Mpc})<40$ scales. For each selection function we derive the expected survey uncertainties $\sigma_i$ on $dn/dz$ and $\xi$
using the $S/N$ calculated from the simulations, described in Appendix ~\ref{appendixB}. 
These account for the integrated effect of the Poisson noise and sample variance.

\subsection{Fiducial cosmology and model parameters}
\begin{figure}
\includegraphics[width=8cm]{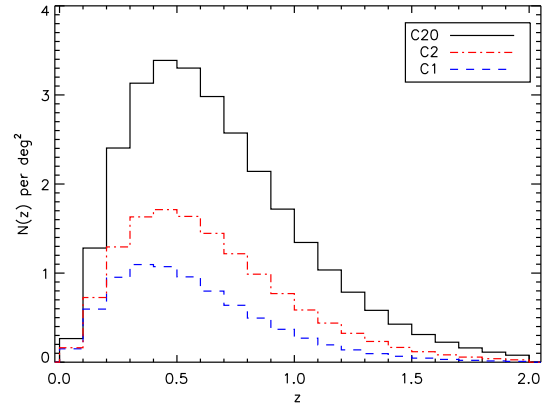}
\caption{Redshift distribution of the C1, C2 and C20 populations for the $\Lambda$CDM fiducial cosmology in 
the {\sl Survey-A} configuration.}
  \label{dndz_fid}
\end{figure}

\begin{figure}
\includegraphics[width=8cm]{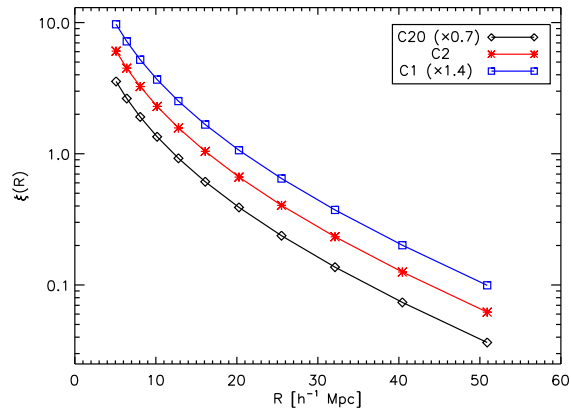}
\caption{Two-point correlation function of the C1, C2 and C20 populations.}
\label{xsi_fid}
\end{figure}

We assume as our fiducial cosmology a flat $\Lambda$CDM model best-fitting the 
WMAP-5 years data \citep{Dunkley09}, specified by the following parameter values: 
$\Omega_{\rm m}h^2=0.1326$, $\Omega_{\rm b}h^2=0.0227$, $h=0.719$, $n_s=0.963$, $\sigma_8=0.796$, $\tau = 0.087$.
For this model the expected number of clusters as function of redshift for {\sl Survey-A} ($50 {\rm deg^2}$)
is shown  in Fig.~\ref{dndz_fid} for the three selection functions.
Fig.~\ref{xsi_fid} displays the $2$-point cluster correlation function. Here it is worth noticing
that while the three functions have the same shape, the C2 curve has a slightly lower amplitude 
than C1, and higher than C20, consistently with the  mass ranges pertaining to these samples (less massive objects are less clustered).

We derive constraints on the following set of parameters: $\Omega_{\rm m},\Omega_{\rm b},h,n_s,\sigma_8$ ($\Lambda$CDM), 
including a
varying equation of state $w(z)=w_0+w_az/(1+z)$ with parameters $w_0$ and $w_a$ \citep{Polarski,Linder} for $w(z)$CDM models.

\subsection{Modeling cluster mass uncertainties in the Fisher analysis}

\begin{figure}
\vspace{3cm}
\vbox{
\includegraphics[width=8cm]{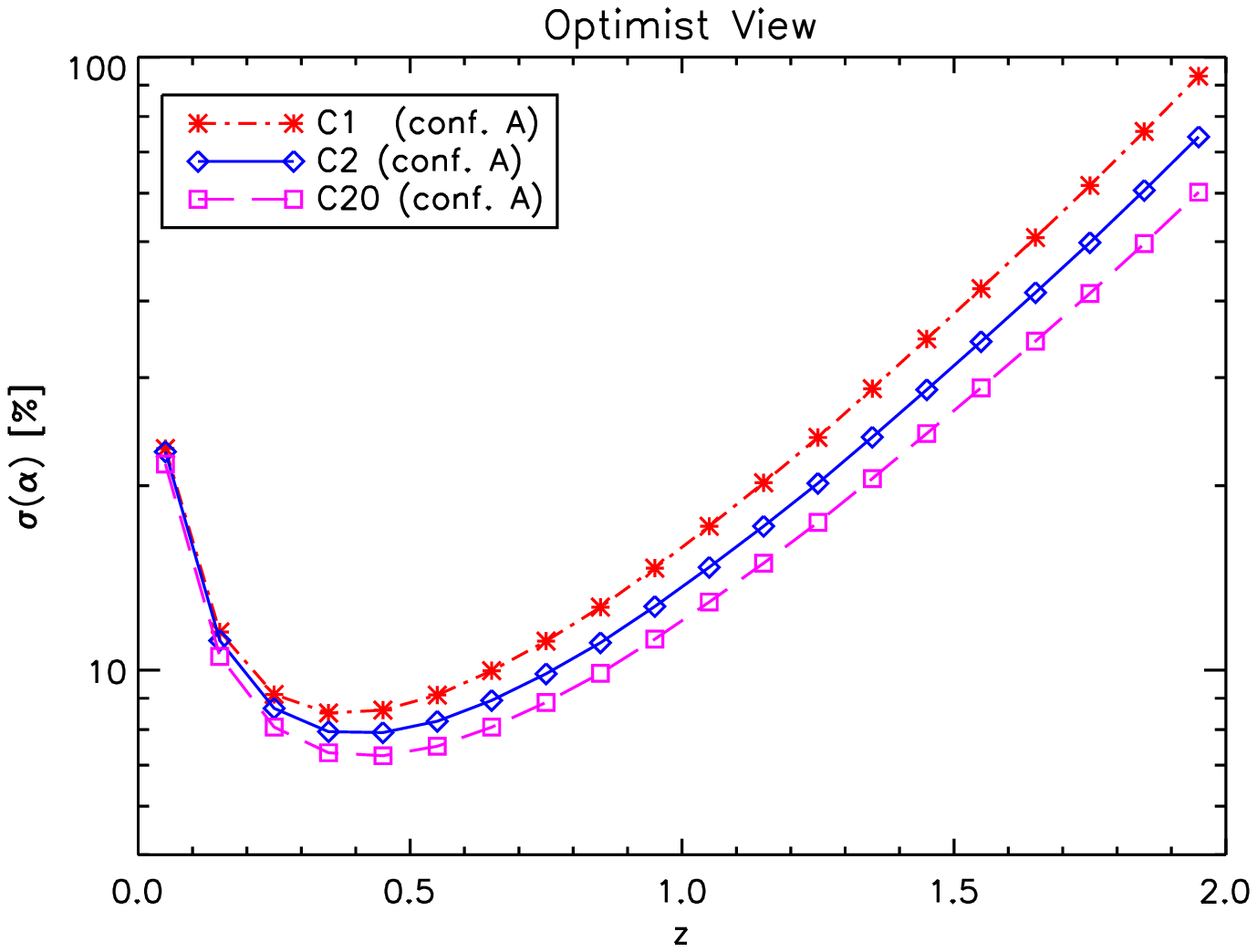}
\includegraphics[width=8cm]{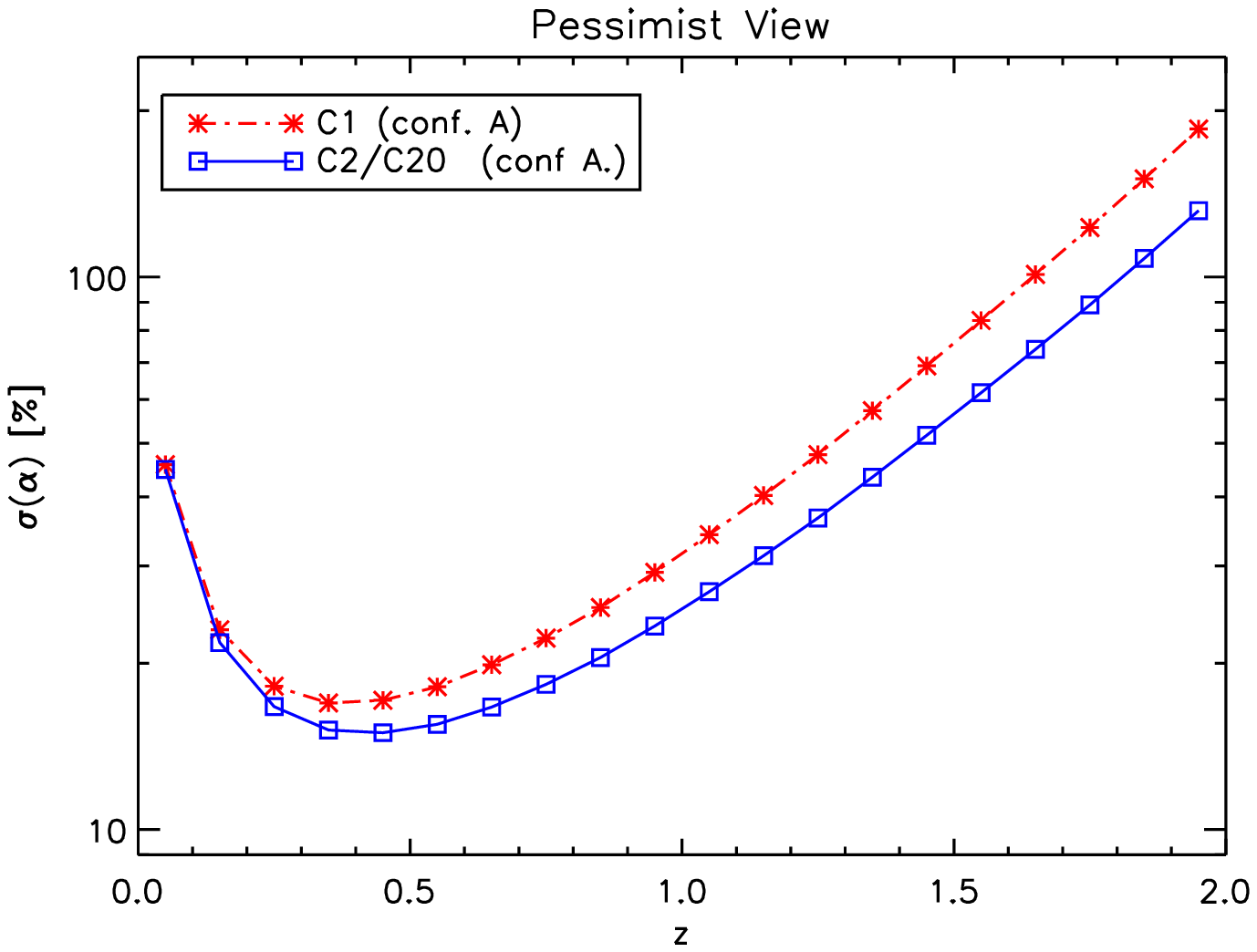}
}
\caption{Priors for the Fisher analysis on the normalization of the M-L relation as a function of redshift, for the optimistic and pessimistic cases for the {\sl Survey-A} configuration}
\label{alphaprior}
\end{figure}

For the Fisher analysis, our aim is to reproduce as much as possible the observational procedure and the subsequent cosmological analysis. To summarize the steps: (1) clusters are selected in the XMM images according to a two-dimensional parameter space; (2) corresponding $dn/dz$ and $\xi$ are derived; (3) each cluster mass is measured at a given accuracy - the mass measurements being cosmology-dependent; (4)  for a given cosmology, we compute $dn/dz$ and $\xi$, the observational selection function being yet translated in the [M,z] space following scaling laws - this is the point where the mass accuracy enters; (5) as already specified, we encapsulate all uncertainties on the scaling laws in the M-L relation for the cosmological modeling; (6) the set of cosmological parameters giving best agreement both on $dn/dz$ and $\xi$,  describes the most likely cosmological model.\\
Practically, in the Fisher analysis, we assume that the slope and the dispersion of the M-L relation are known and do not depend on redshift. We let, however, the normalization of the relation free as a scale factor $\alpha(z)$. We take one scale factor for each  redshift bin ($\Delta (z)=0.1$), hence we have 10 or 20 nuisance parameters depending on the survey depth.
The priors for the analysis are derived from the accuracy assumed for the mass measurements of the individual clusters (Table \ref{masserr}); they are displayed in Fig. \ref{alphaprior} for the optimistic and pessimistic cases.

\subsection{Planck Fisher Matrix}


To estimate the full cosmological yield of an XXL-survey, we perform a
 joint analysis of the cluster survey with
the primary CMB power spectra (temperature-TT, polarization-EE 
and cross-correlation TE) soon to be measured by the Planck satellite. 

A precise assessment of the Planck capabilities would require to model
in detail the map making and component separation processes. 
To circumvent this problem, we make the simplifying assumption that the 
sky images in the three bands where the CMB emission dominates (100, 143 and 
217~GHz) are readily usable to measure the power spectra, while the other 
bands permit a perfect characterization of the other contaminating signals.

Following \citep{Zaldarriaga97}, the noise covariance matrix
for each $l$ (including the cosmic variance) is then given by: 

\begin{eqnarray}
{\rm Cov}(C_l^{TT},C_l^{TT})&=&\frac{2}{(2l+1)f_{\rm sky}}(C_{l}^{TT}+N_{l,TT}^{-2}),\nonumber\\
{\rm Cov}(C_l^{EE},C_l^{EE})&=&\frac{2}{(2l+1)f_{\rm sky}}(C_{l}^{EE}+N_{l,EE}^{-2}),\nonumber\\
{\rm Cov}(C_l^{TE},C_l^{TE})&=&\frac{1}{(2l+1)f_{\rm sky}}[C_{l,TE}^{2}\nonumber\\
&+&(C_{l}^{TT}+N_{l,TT}^{-2})(C_{l}^{EE}+N_{l,EE}^{-2})],\nonumber\\
{\rm Cov}(C_l^{EE},C_l^{TE})&=&\frac{2}{(2l+1)f_{\rm sky}}C_l^{TE}(C_{l}^{EE}+N_{l,EE}^{-2})\nonumber\\
{\rm Cov}(C_l^{TT},C_l^{TE})&=&\frac{2}{(2l+1)f_{\rm sky}}C_l^{TE}(C_{l}^{TT}+N_{l,TT}^{-2})\nonumber\\
{\rm Cov}(C_l^{TT},C_l^{EE})&=&\frac{2}{(2l+1)f_{\rm sky}}C_{l,TE}^2\nonumber,\\
\end{eqnarray}
where
\begin{equation}
N_{l,X}^2=\sum_c(\sigma_{c,X}\phi_{c})^{-2}e^{-l(l+1)\phi_{c}^2/(8\log{2})},
\end{equation}
is the contribution of the instrumental noise to the uncertainty on the spectrum $X$, which results 
from averaging over the different frequency channels $c$, with sensitivity $\sigma_{c,X}$ and angular 
bean-width $\phi_{c}$. 
In Table~\ref{wmaplanck}, we quote the assumed experimental characteristics for the Planck satellite, which 
we obtained from the mission definition document (the so-called `{\it Bluebook}')\footnote{available 
from the ESA web pages of the Planck mission: http://www.rssd.esa.int/index.php?project=Planck}. 
We adopt a fractional sky coverage of $f_{\rm sky}=0.8$ to account for the masking of the galactic plane.

\begin{table}\caption{Planck survey parameters.}\centering
\begin{tabular}{lccc}
\hline\hline
{}&Planck&{}&\\
\hline\hline
{\rm Frequency (GHz)}& 100 & 143 & 217\\
$\phi_c~({\rm arcmin})$&10.0& 7.1&5.0\\
$\sigma_{c,T}~({\rm \mu K})$& 6.8& 6.0 & 13.1\\
$\sigma_{c,E}~({\rm \mu K})$&10.9& 11.4 & 26.7\\
\hline \hline
\end{tabular}\label{wmaplanck}
\end{table}

The full CMB Fisher matrix for a set of cosmological parameters $(\theta_\mu)$
is straightforwardly obtained as:
\begin{equation} 
F_{\mu\nu}^{\rm CMB}=\sum_{l}\sum_{X,Y}\frac{\partial{C}_{l}^{X}}{\partial\theta_\mu}{\rm Cov^{-1}}(C_l^X,C_l^Y)\frac{\partial{C}_{l}^{Y}}{\partial\theta_\nu},
\label{fisherCMB}
\end{equation}
where $X,Y=TT,EE,TE$ and we sum over $l$ values in the range [1,2000].

In practice, we compute the power spectra using the CMBFAST code and some care has to 
be taken in order to correctly account for the intrinsic CMB degeneracies.
Indeed, the shape of the matter power spectrum at the recombination epoch
is only a function of the primordial power spectrum and the physical densities 
($\rho_{\rm m}$, $\rho_{\rm b}$, $\rho_{\rm r}$) in the early universe. 
Further, while the relative amplitudes of the CMB peaks depend on the details of 
the matter/photon densities, the physical scale of the baryon oscillation pattern 
is simply proportional to the sound horizon at recombination ($r_{\rm s}$). 
As a consequence, the CMB observables only depend on $h$, $\Omega_{\rm DE}$, $w_0$ and 
$w_a$ through the so-called CMB acoustic scale:
\begin{equation}
  l_{\rm a} = \pi (1+z_{\rm dec}) \frac{d_{\rm a}(z_{\rm dec})}{r_{\rm s}}
\end{equation}
where $z_{\rm dec}$ is the redshift of decoupling and $d_{\rm a}$ the angular diameter distance. 
[the factor $(1+z_{\rm dec})$ comes from the fact that $r_{\rm s}$ is measured in the comoving
frame]. This exact degeneracy of the CMB, known as the geometric degeneracy, prevents
CMB experiments from giving any constraint on the dark energy without adding other 
observables.
Numerical estimates of CMB Fisher matrices, based on codes such as CMBFAST, fail at 
accurately reproducing this degeneracy (see e.g. \citet{Kosowsky02}) and tend to give 
unrealistic results solely because of numerical uncertainties. 
We therefore follow the approach of the DETF report and Rassat et al. (2009) to estimate 
the Fisher matrix over a `natural' set of cosmological parameters ($\Omega_{\rm m} h^2$, $\Omega_{\rm b} h^2$, 
$l_{\rm a}$, $\Delta_{\mathcal{R}}^2$, $n_s$ and $\tau$ ). 
We then marginalize over $\tau$ and convert the Fisher matrix into our preferred parameter set 
using the Jacobian matrix of the transformation.\\
In Table 4 we quote the resulting constraints for Planck alone with
or without the use of polarization.
Because of the geometrical degeneracy, only constraints on the
simplest ${\Lambda}$CMD models can be obtained, however
we have also estimated the full Fisher Matrix for the w(z)CDM
model, since it is necessary to derive the combined constraints from
the Planck CMB spectra with the cluster observables.
This is then simply achieved by adding Eq.~(\ref{fisherCMB}) to Eq.~(\ref{fisher}).

\begin{table}\centering
\caption{Fisher matrix errors on the cosmological parameters from Planck.}
\begin{tabular}{lcccc}
\hline
\hline
 & \multicolumn{2}{c}{10\% prior on $h$} & \multicolumn{2}{c}{Flat universe} \\
\hline
 & TT & TT+TE+EE & TT & TT+TE+EE\\
\hline
\hline
$h             $  &     0.0719  &     0.0719  &     0.0030  &     0.0017  \\
$\Omega_{\rm b}      $  &     0.0088  &     0.0088  &     0.0007  &     0.0005  \\
$\Omega_{\rm m}      $  &     0.0514  &     0.0513  &     0.0019  &     0.0010  \\
$\Omega_\Lambda$  &     0.0880  &     0.0879  &      -      &      -      \\
$\sigma_8      $  &     0.0536  &     0.0361  &     0.0400  &     0.0067  \\
$n_s           $  &     0.0070  &     0.0040  &     0.0070  &     0.0040  \\
$\tau          $  &     0.0532  &     0.0040  &     0.0532  &     0.0040  \\
\hline
\hline
\end{tabular}
\label{wacmb}
\end{table}

 \section{ Predicted constraints on the cosmological parameters} 
 
 \label{constraints}

\begin{figure*}
\hbox{ \includegraphics[width=8cm]{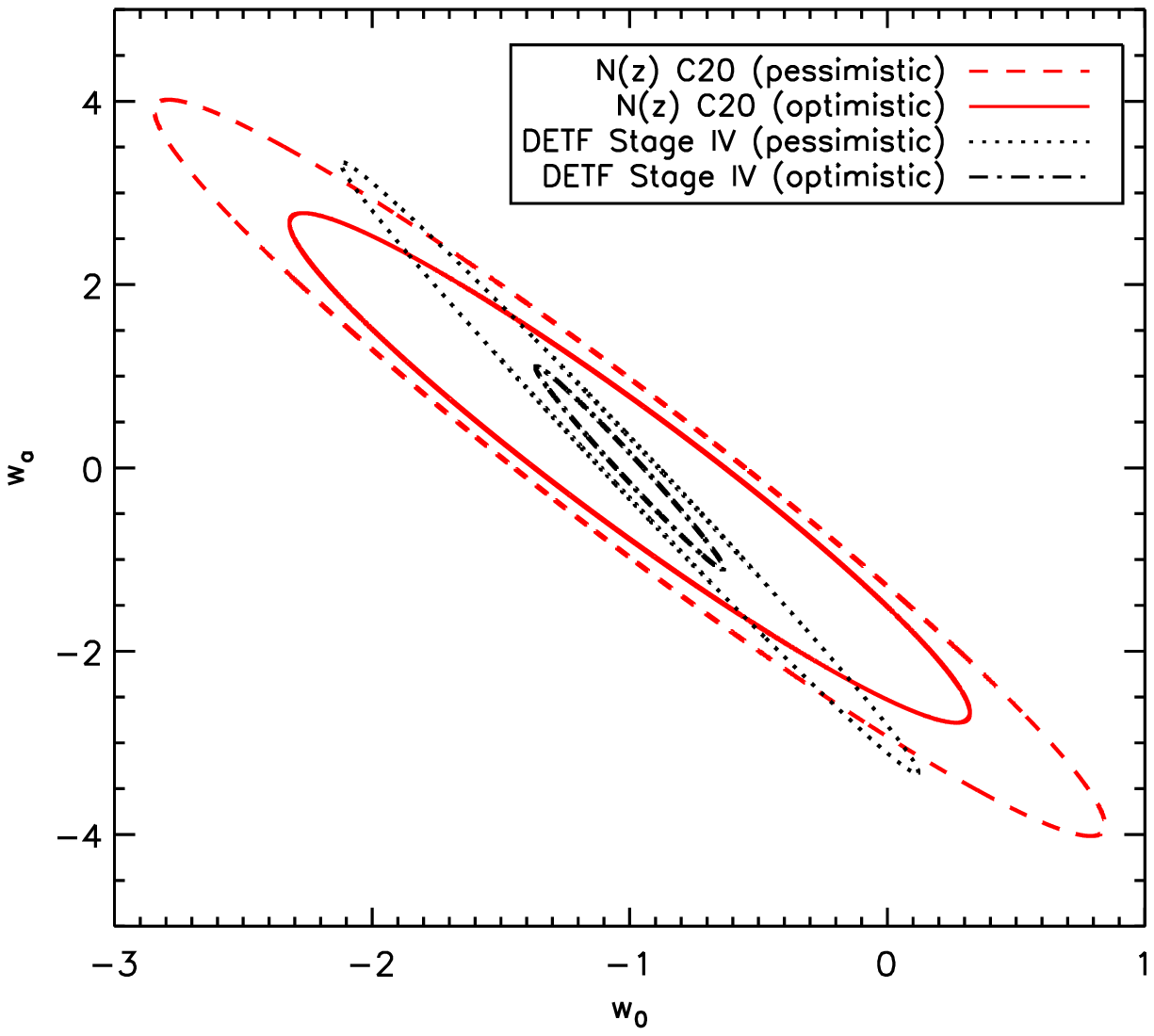} \includegraphics[width=8cm]{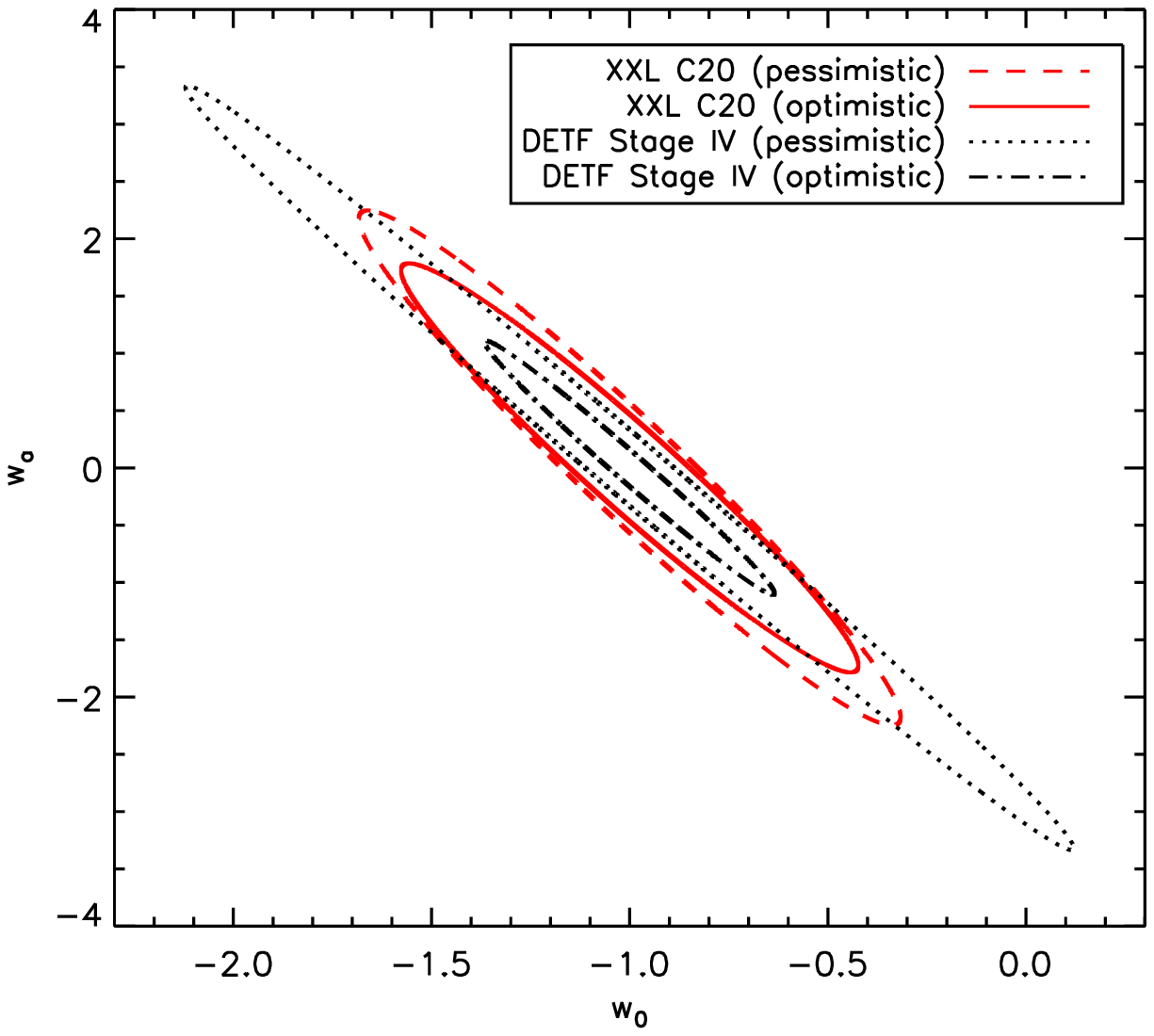} }
\caption{Illustration of the impact of the cluster-cluster correlation function - {\sl A2 Survey} configuration (from Table \ref{fisherA2-40}). {\em Left}: 1-$\sigma$ cosmological constraints from the cluster number counts only.   {\em Right} Adding the correlation function.  The black contours indicate the predictions by Dark Energy Task Force, stage IV . }
  \label{effet-xsi}
\end{figure*}

Results from the Fisher analysis for the equation of state of the dark energy are presented in Tables \ref{fisherA2-40} and \ref{fisherB0} for the A2 and B0 survey configurations.  We display the ultimate accuracy
that can be reached for the most general, non flat, w(z)CDM  cosmology.  We outline below the main outcome of the study.\\
(1) The comparison between the C1 and C2 populations (limited to $0<z<1$) shows an improvement on $w_{0}$, $w_{a}$ of about 20, 10\% for the C2 sample. The C2 clusters are roughly twice as numerous as the C1, but less massive in average so that their impact on cosmological measurements is expected to be indeed relatively smaller.\\
(2) Focussing on the {\sl Survey-A} configuration,  the C20 clusters are four times more numerous than the C1 and some 250 of them are between $1<z<2$. The net effect is an improvement better than a factor of two  on $w_{a}$ and $w_{0}$.
\\(3) The comparison between the B and A survey designs  for the C2 and C20 populations respectively shows comparable constraints when $dn/dz, ~ \xi$ and Planck are combined (optimistic and pessimistic cases). However, the total number of clusters involved is 2320 for B compared to only  1400 for A. This stresses the efficiency of the $1<z<2$ clusters for characterising the dark energy (see also \citet{Baldi10}).\\
(4) Table  \ref{fisherA2-10} lists the constraints expected after the first scan of survey A, thus at 1/4 of its nominal depth (C2 population only and measured in pessimistic conditions): the accuracy is about half of that at full depth, hence along the line of the $\rm signal \propto \sqrt{\rm time}$ ratio.\\ 
We have further investigated the role of various hypotheses that were made in the prescription of the Fisher analysis.\\
(5) This study is amongst the first ones to qualitatively consider  the added value of the cluster spatial distribution  in the determination of the DE parameters  \citep[see also][]{Mohr04, Huetsi10}. The impact of $\xi$ is highlighted in Fig. \ref{effet-xsi}.
It is remarkable   given that the regions considered for {\sl Survey-A2} are 
only 3.5 deg aside, but should not be considered as unexpected. In fact,
$\xi$ is particularly sensitive to $\Omega_{\rm m}$ and $\sigma_8$, thus it 
strongly contributes to breaking model parameter degeneracies. Furthermore 
the mass dependence of the halo clustering is opposite to that of the 
number counts.  On the one hand, less massive 
halos are less clustered than the massive ones; on the other hand, the 
former are more numerous. Thus a combined measurement allows for a 
better mass determination of the cluster sample and directly improves the 
parameter inference. This is a clear advantage of dedicated 
cluster surveys over serendipitous searches.\\
(6) Introducing a prior of 10\% on the Hubble constant  does not significantly improve  $w_{a}$, $w_{0}$ for the final $dn/dz + \xi$ + Planck settings but some 40, 20\% better constraints are predicted when only $dn/dz$ + Planck are considered.\\   
(7) We have examined the case where the M-L relation is perfectly known at all redshifts: we observe an improvement of less than 15\% both on $w_{0}$ and $w_{a}$ for the C20 population with the optimistic assumption.\\  
(8) We have further investigated what happens if the dispersion in the M-L relation (which can be interpreted as the dispersion in any mass-observable relation) is decreased from 0.6 to 0.1, re-computing the priors accordingly.  In this case, the improvement is  $\sim$ 10\%; assuming in addition that the M-L relation is perfectly known leads to a negligible improvement.\\ 
(9)  We have assumed that the cluster luminosities evolve self-similarly,  which tends to be supported by current observations \citep{Maughan08}. Other scaling laws can be assumed like,  for instance, no evolution, which implies that distant clusters are less luminous than in the self-similar hypothesis: this would decrease the number of detected high-z clusters. The impact of the cluster evolution hypothesis can be  bracketed by the extreme case were no $z>1$ clusters are detected; in this case, the optimistic constraints on $w_{0}$, $w_{a}$ would change from  0.40, 1.29 to 0.51, 1.67.\\ 
(10) Finally, assuming a flat w(z)CDM cosmology improves the determination of $w_{0}$ and $w_{a}$ by about 5\%. For a flat $w$CDM, we predict a precision of  0.040 for $w$ with the C20 optimistic configuration ({\sl Survey-A}) .\\

\begin{figure*}
\hbox{ \includegraphics[width=8cm]{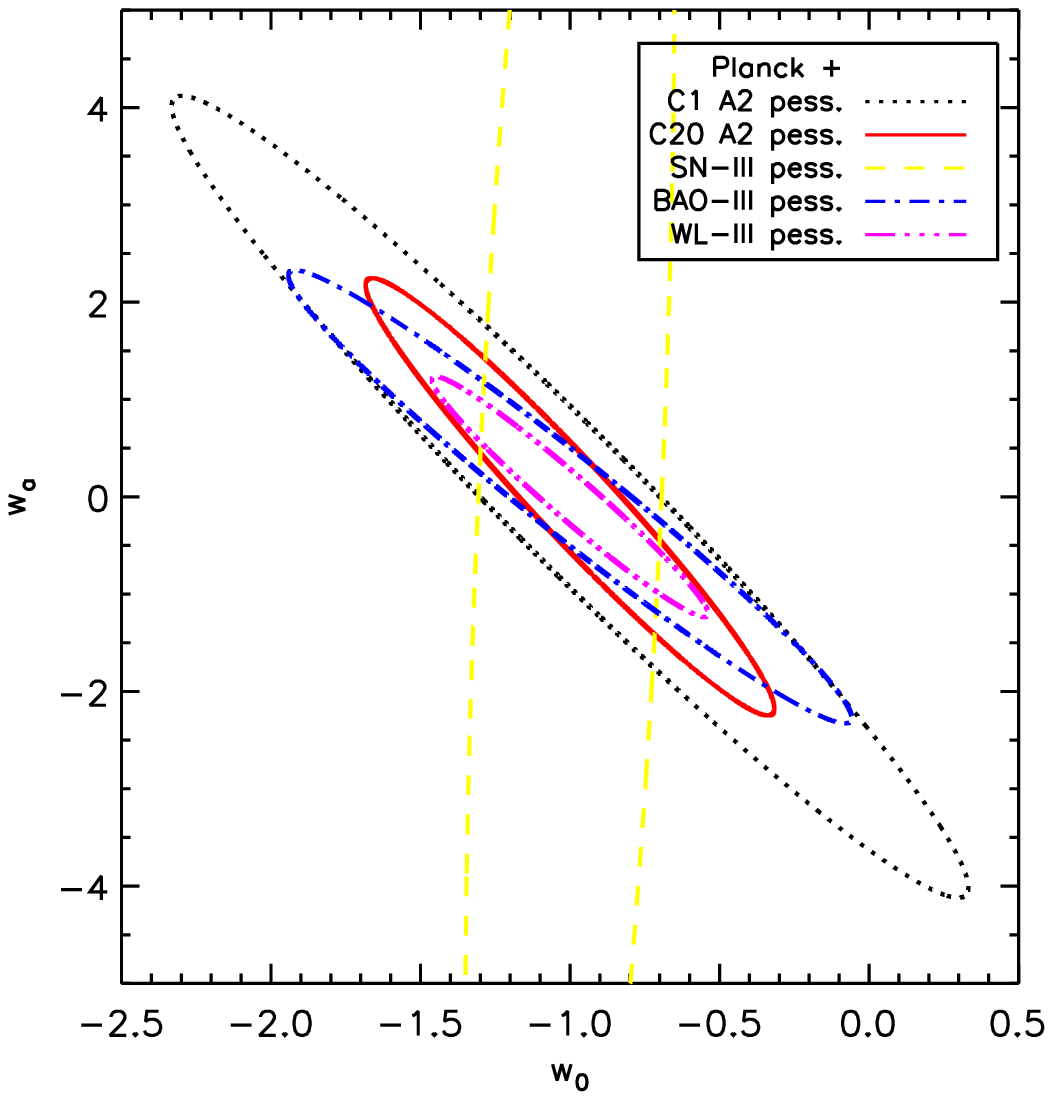} \includegraphics[width=8cm]{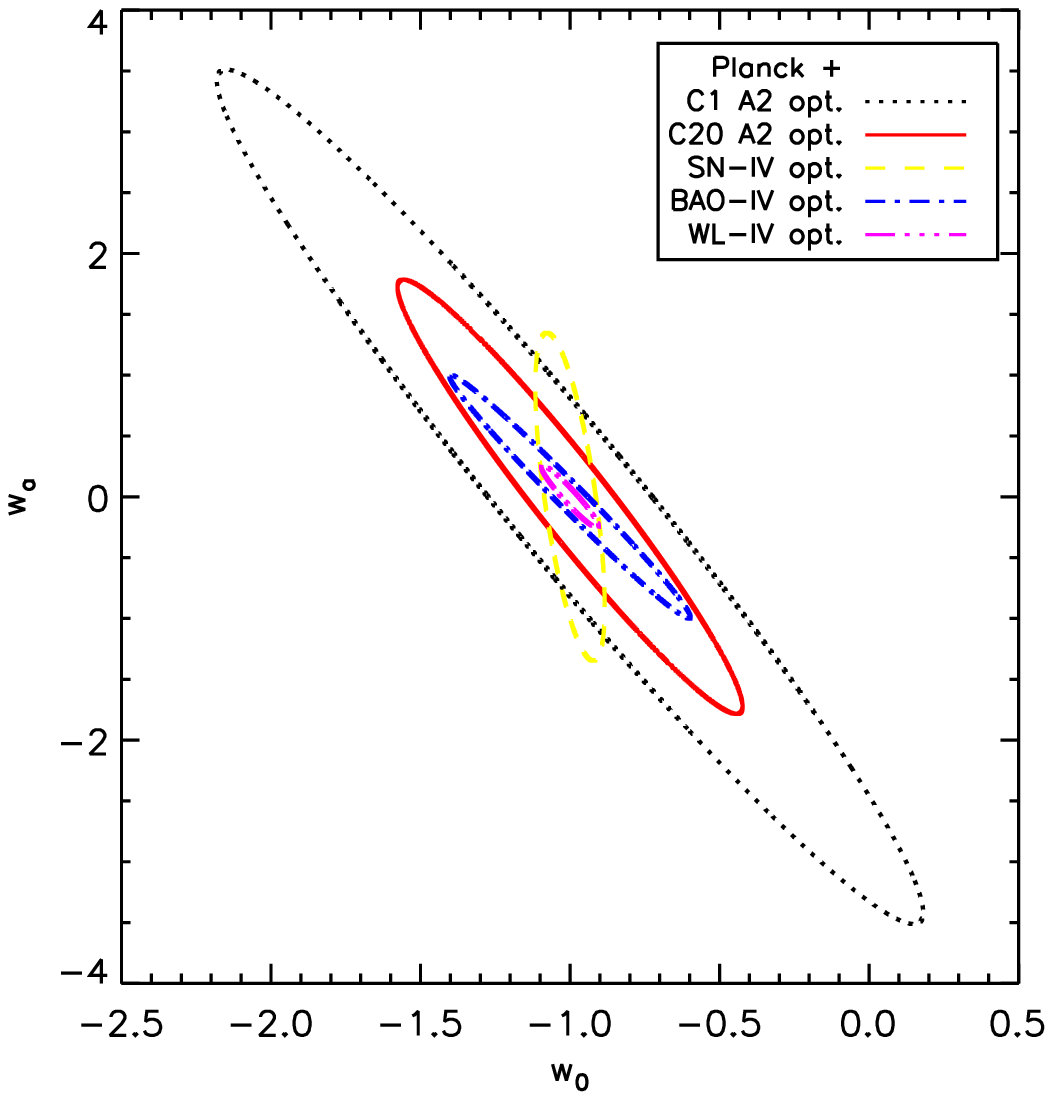}}
\caption{{\em Left}: the 1-$\sigma$ contours in the $w_{0}$-$w_{a}$ plane for the various DETF probes at stage III (pessimistic assumptions) along with the XXL predictions for the C1 and C20 populations (pessimistic case from Table \ref{fisherA2-40}). The contours for the DETF probes have been derived using the Fisher matrix data from the DETFast code, including the Planck priors in the same way as for XXL. {\em Right}, same as {\em Left} for the DETF stage IV and optimistic assumptions. }
  \label{detf-compare}
\end{figure*}

A general summary of the expected dark energy parameter uncertainties from future cluster surveys has been presented in the
Dark Energy Task Force (DETF) document \citep{Albrecht06}. This review study classifies the projected performances of cluster surveys 
into stage II, III and IV. Stage II corresponds to surveys of $200~{\rm deg^2}$ with a mean mass threshold of $10^{14} h^{-1}{\rm M_\odot}$
detecting approximately $4000-5000$ clusters, and for which the expected errors on the dark energy 
parameters are $\sigma_{w_0}=1.1$ and $\sigma_{w_a}=3.2$. Stage III consists of surveys covering $4000~{\rm deg^2}$ 
with a mean threshold of $10^{14.2} {h^{-1} \rm M_\odot}$ detecting $\sim30,000$ clusters.
 Finally Stage IV corresponds to surveys covering $20,000~{\rm deg^2}$ 
with a mass threshold of $10^{14.4}{h^{-1}\rm M_\odot}$ and providing also $30,000$ clusters.The DETF predictions for stage III and IV are recalled in Table \ref{fisherdetf}; they are comparable for both stages as each of them appear to be dominated by systematics.\\
These projections have been derived under
a number of assumptions which differ from ours. First, the halo mass function has been assumed in the fitting form provided by \citet{Jenkins01}.
Second, the settings of the Fisher analysis are also slightly different: while both studies involve the same number of parameters, the analysis presented by \citet{Albrecht06}  assumes a prior of $\sim 10\%$ on the Hubble constant - we do not (they also consider the $\delta_{\zeta}$ parameter ($k^{3}P_{\zeta}/2\pi^{2}$) in place of $\sigma_{8}$). Conversely, they use only number counts - we consider, in addition, the correlation function. The DETF adopts a constant mass selection, and masses are supposedly determined through  ``self-calibration'', i.e. a functional dependence between flux 
(or richness), mass and redshift is assumed \citep[see][]{Mohr04}. The DETF has further assumed
a root mean square error in the mean/variance of mass per redshift bin ranging from $2-14\%$ for 
stage III and $1.6-11\%$ for stage IV.  Despite these differences, it is worth comparing the performances advocated by the DETF with our predictions. A quick glance at Tables   \ref{fisherA2-40} and \ref{fisherdetf} immediately reveals that the XXL pessimistic predictions outperform the DETF pessimistic ones and that XXL optimistic ones lay between the optimistic and pessimistic DETF calculations. This is a somewhat unexpected result given the ratio of the surveyed areas (a factor of 80-400) but is readily understandable as the effect of the mass accuracy and of the presence of $z>1$ clusters, a direct consequence of the XMM deep exposures. We further compare the virtue of the XXL cluster population with the other cosmological probes examined by the DETF, namely :  baryon acoustic oscillations, supernovae and  weak lensing measurements.  The comparisons are displayed in Fig. \ref{detf-compare}.

\section{Discussion and conclusions}

\label{conclusion}

We have shown through a Fisher matrix calculation that the XXL {\sl Survey-A} (and {\sl B}) can provide measurements of the cluster number counts and $2$-point correlation function of sufficient precision to provide useful constraints on the equation of state of the dark energy.
In our analysis, special care has been devoted to the realistic modeling  of the statistical uncertainties (sample variance and shot noise) due to the small size of the surveyed area (50 or 200 \dd) and of the cluster mass measurements. Our experience gained with XMM has allowed us to consider realistic cluster selection functions and to apply priors on individual cluster mass measurements. We have favored this approach against the use of  `self-calibration' techniques, intended to by-pass the current ignorance about the evolution of  the cluster scaling law by simultaneously fitting its functional form with cosmology.
After all, self-calibration has not been observationally tested yet, 
and as shown by \citet{Sahlen09}, it is hampered by the fact that it introduces 
a latent degeneracy between the dispersion in the scaling laws and their redshift evolution. Moreover, it has been
pointed out by \citet{pacaud07} that, in the case of X-ray flux measurements, 
emission lines produce discontinuities which cannot be simply accounted for 
by the parametrized functional dependence. On the basis of these
considerations we have deliberately assumed to individually measure cluster masses, and improve 
the X-ray mass derivation by means of S-Z and weak lensing observations.
This is a reasonable working assumption given the relatively limited size of the surveyed area and the results  will form the ideal basis for investigating, a posteriori,  self-calibration techniques.
In the present analysis we have let the normalization of the scaling relation to be free for each $\Delta z=0.1$ redshift bins. Alternatively, taking larger bins (e.g. $\Delta z=0.2$) decreases the number of free parameters by a factor of two. This would allow the introduction of e.g. two more free parameters so as to enable the simultaneous fit of  the evolution of the slope and of the dispersion of the relation. These hypotheses will be discussed in a subsequent article (paper II, Pacaud et al, in prep.).
In this forthcoming work, we shall also compare the relative efficiency of various cluster selection functions (such as those presented here and a fixed mass limit at any redshift), investigate the role of plausible evolution laws other than self-similarity, examine the impact of the DE inhomogeneities on the halo mass function, discuss the added value of the evolution of $\xi$  and, especially, that of the cluster mass function ($dn/dMdz$) in constraining the DE equation of state. \\
In any case, our analysis demonstrates that  a medium deep 50 \dd\ survey with XMM - a modest project compared to the DETF stage IV requirements - is in a position to fulfill competitive  expectations in terms of cluster cosmological  studies, while providing constraints which are complementary to those expected from other probes.  Moreover, from a practical point of view, compared to the cluster surveys advocated by the DETF (Stage III and IV) the XXL survey contains some 20 times less clusters, which makes the sample much more tractable.
 
We have shown that the {\sl Survey-A}  and {\sl Survey-B} configurations provide equivalent constraints on the DE
for a similar amount of XMM observing time ($\sim 20$ Ms). Practically, we favor configuration A over B as, besides constraining the properties of dark energy, it is observationally more advantageous. There are also a number of compelling arguments 
as to the ``legacy value'' of  {\sl Survey-A}, which make it more appealing. Let us review them in some detail.\\
- The aimed mass accuracy   (to be complemented by a joint analysis of S-Z and weak lensing surveys), for all clusters entering the analysis, will 
have an invaluable scientific potential for the study of baryon physics. In particular, it will provide the long expected scaling law evolution 
out to a redshift of $\sim1.5$ and to a mass $M_{200} \sim 10^{14} M_{\odot}$. XMM pointed observations cannot achieve such an efficient determination 
for the simple reason that few X-ray clusters, and only massive ones,  are known at $z\sim1$. In contrast the {\sl Survey-A} configuration has the  ability to detect and reliably measure the signal from these objects in one single shot. This will provide very useful calibration data for other 
surveys (e.g. DES, eRosita), which are expected to cover much larger areas but at lower depth and poorer X-ray angular resolution \citep{Predehl06}. Then the self-calibration method will be easily testable.\\
- The spatial distribution of X-ray AGNs, which will constitute more than 90\% of the sources of the planed survey, will be studied on very 
large scales as a function of their spectral properties. \\
- For visibility reasons and observation programming, we favor the splitting of {\sl Survey-A} in two or four sub-regions spread in right ascension. Furthermore, the XMM observations can be scheduled over four years, with each field being entirely covered by 10 ks XMM 
observations every year. The first year scan could already provide the full C1 + C2 cluster catalogue, hence measurements of 
$\xi$ and $dn/dz$ and constraints on the DE to an accuracy half of the final value. The three subsequent scans will then increase the number of X-ray photons down to the nominal 40 ks depth, 
thus providing the spectral accuracy and, finally, the cluster mass accuracy required for the full cosmological analysis.\\
- Spreading the XMM observations over four years can provide unrivaled information about AGN variability over 
large timescales as a function of the spectral properties and environment.\\

Finally, in addition to the important added value of $\xi$, we mention a number of arguments leading to favor contiguous surveys with respect to serendipitous cluster searches:\\
- Operationally it is much more efficient to perform a joint X-ray + optical/lensing + S-Z survey than to undertake 
a pointed follow-up of X-ray clusters. And obviously, a joint optical survey renders the X-ray source identification straightforward.  \\  
-  Homogeneous wide surveys, compared to serendipitous searches,
highly simplify the derivation of the selection functions which,
as shown here, play a critical role for cosmological studies. \\
- Using XMM archival data would only allow the determination of $dn/dz$ and it is moreover important to note that the situation is different from that of the ROSAT serendipitous searches.  ROSAT had a two-degree diameter field of view (against 30 arcmin for XMM) and a significant fraction of the known cluster population has been imaged by XMM\footnote{Over 1 000 cluster observations performed. Out of the some 1600 observations available with exposure time longer than 40ks (any type of targets), about 400 useful images remain when considering only high galactic latitude public observations and assuming a flaring rate of 25\% (status of the XMM archive by  September 2010)}. This introduces complex biases that cannot be removed by simply discarding the central target or ignoring the target clusters as was routinely assumed in the past; it is especially serious at high redshift since only the X-ray brightest known clusters were considered as targets. \\
- The proposed homogeneous survey will also enable the determination of the structure of the X-ray 
background on very large scales at energies ranging from 0.1 to 10 keV. In addition, once the cluster population is detected and 
the redshifts measured, their 3-D distribution will enable the identification of putative cosmic filaments. 
Staking the X-ray data corresponding to the location of many filaments then could lead to the first detection of the Warm Hot 
Intergalactic Medium in emission \citep{Soltan08}. \\

One of the interesting outcomes of the present study is to have quantitatively estimated the impact of the cluster-cluster correlation function in dark energy studies.
We leave to future studies the possibility of measuring the evolution of
the cluster mass function $dn/dMdz$ rather than $dn/dz$ with the
XXL survey, as well as the combination with the low-z REFLEX correlation function and the Planck cluster number counts +
correlation function. 
In the future, one can also well imagine constraining cosmology directly by applying
the X-ray selection function on a large set of hydrodynamical
simulations - when these become achievable - 
and match the properties of the resulting simulated cluster catalogues
to that of the observed XXL one. Such methods, which are already applied on the Ly$\alpha$ forest \citep{Viel06} would allow one to totally
by-pass the determination of the cluster mass-observable relations as
a function of redshift.

\section*{Acknowledgments}
We acknowledge numerous discussions which took place during the XXL meeting, held in Paris in April 2008 and from which the XXL survey concept emerged.  We would like to thank the authors of Pinocchio, especially
P. L. Monaco, for their help and support on their software. We thank  Romain Teyssier for having provided us with the Horizon cluster catalogue. We are grateful to Ana\"is Rassat  and Alexandre Refregier  for many useful technical discussions and to Jim Rich for helpful comments on the manuscript. Part of this work is financed by a grant from the Centre National d'Etudes Spatiales. FP acknowledges support from the Transregio Programme TR33 of the Deutsche Forschungsgemeinschaft and from the grant 50\,OR\,1003 of the Deutsches Zentrum f\"ur Luft- und Raumfahrt. 

 \onecolumn

\begin{table*} \centering
\caption{Cosmological constraints. Survey configuration A2 - 50 \dd\ full depth (40 ks XMM exposures)~~~~~~ 1-$\sigma$ errors on $w_{0}$   /  $w_{a}$}
\begin{tabular}{llccccc}
\hline \hline
& &  \multicolumn{2}{c}{Pessimistic mass measurements}  & \multicolumn{2}{c}{Optimistic mass measurements} \\
Selection & Redshift range &  dn/dz + Planck & dn/dz + $\xi$ + Planck & dn/dz + Planck & dn/dz + $\xi$ + Planck\\
\hline\hline
 C1 & $0<z<1 $  &2.38 / 5.08 &0.88 / 2.71 &1.98 / 4.15 &0.78 / 2.32 \\
C2  & $0<z<1 $  &2.00 / 4.64 & 0.72 / 2.36 &1.70 / 3. 89 & 0.65 / 2.06\\
C20 & $0<z<2 $  & 1.19 / 2.59  & 0.45 / 1.46 & 0.87 / 1.82 & {\bf 0.38 / 1.18} \\
\hline \hline
\end{tabular}
\label{fisherA2-40}
\end{table*}
 
 \begin{table*}\centering
\caption{Cosmological constraints. Survey configuration B0 - 200 \dd\ full depth (10 ks XMM exposures)~~~~~~ 1-$\sigma$ errors on $w_{0}$   /  $w_{a}$}
\begin{tabular}{llccccc}
\hline \hline
& & \multicolumn{2}{c}{Pessimistic mass measurements}  & \multicolumn{2}{c}{Optimistic mass measurements} \\
Selection & Redshift range & dn/dz + Planck & dn/dz + $\xi$ + Planck & dn/dz + Planck & dn/dz + $\xi$ + Planck\\
\hline\hline
 C1 & $0<z<1 $ & 1.58 / 3.30 & 0.54 / 1.71 & 1.33 / 2.72& 0.48 / 1.47 \\
C2  & $0<z<1 $  & 1.42 / 3.29 & 0.47 / 1.60 & 1.13 / 2.52 & {\bf 0.40 / 1.29} \\
\hline \hline
\end{tabular}
\label{fisherB0}
\end{table*}

 \begin{table}\centering
\caption{Cosmological constraints. Survey configuration A2 - 50 \dd\ 1/4 depth (10 ks XMM exposures)~~~~~~ 1-$\sigma$ errors on $w_{0}$   /  $w_{a}$}
\begin{tabular}{llcc}

\hline \hline
Selection & Redshift range & dn/dz + Planck & dn/dz + $\xi$ + Planck  \\  \hline\hline
C1 (pessimistic)  & $0<z<1 $ &  2.77 / 5.98 &  {\bf 0.97 / 3.08}\\
C2 (optimistic)  & $0<z<2 $ &  1.14 / 2.44 & {\bf 0.55 / 1.70}\\
\hline \hline
\end{tabular}
\label{fisherA2-10}
\end{table}

\begin{table}\centering
\caption{Cosmological constraints from clusters following the DETF survey designs ~~~~~~ 1-$\sigma$ errors on $w_{0}$   /  $w_{a}$}
\begin{tabular}{lcc}
\hline \hline
Stage  & Pessimistic  & Optimistic  \\
\hline\hline
III  & 0.70 / 2.11 & 0.26 / 0.77 \\
IV  &  0.73 / 2.18 & 0.24 / 0.73 \\
\hline \hline
\end{tabular}
\label{fisherdetf}
\end{table}

 \twocolumn

\appendix

\section{Adopted halo mass function}

\label{appendixA}

Early modeling of the mass function relied
on semi-analytical approaches \citep{PressSchec74,Bond91}, however comparison with N-body 
simulations showed discrepancies with the numerically estimated function, and 
a simulation calibrated formula was proposed by \cite{ShethTor99}. 
Over the years the increasing resolution of numerical simulations has led 
to more accurate estimations of the halo mass function, and the standard of accuracy 
has been set by the analysis of \cite{Jenkins01}. The authors of this study have provided 
a `universal' (hence applicable to different cosmologies and at different redshifts) 
fitting formula that is accurate to within $20\%$. Recent studies
have cast doubts on the universality of the mass function. In
particular the analysis by \cite{Tinker08} has shown important
deviations in the high mass end and at high redshift. Nonetheless 
these authors have been able to provide a fitting formula accurate to
$<5\%$ at $z=0$ and to $<20\%$ at $z=1.25$, while degrading
to $50\%$ only at $z=2.5$. In our analysis we assume their fitting halo mass function, 
parametrized in terms of the halo mass enclosed in a radius containing $200$ 
times the critical density of matter, $\mathbf{M_{200c}}$, with the following functional form,
\begin{equation}
f(\sigma,z)=A\left[\left(\frac{\sigma}{b}\right)^{-a}+1\right]e^{-c/\sigma^2},
\end{equation}
where $A=A_0(1+z)^{-0.14}$, $a=a_0(1+z)^{-0.06}$, $b=b_0(1+z)^{-\alpha}$ and $\log_{10}\alpha=-\left[0.75/\log_{10}{(2.67/\Omega_{\rm m}(z))}\right]^{1.2}$,
\citep[see Eqs.~(3)-(8) in][]{Tinker08}. In Table 2 of the same paper, values of the parameters $A_0$, 
$a_0$, $b_0$ and $c$ are provided for several density contrasts $\Delta_{\rm m}$, defined with respect to 
the mean matter density. Following the guidelines of their Appendix B, we perform spline interpolation
between the individual parameter values to match our mass overdensity convention $\Delta_{\rm m} = 200/\Omega_{\rm m}(z)$ at any given z.
This ensures that the mass definition of our cosmological modeling matches the convention used for cluster
scaling relations and thus for our selection function.\\
It has recently been pointed out that DE leaves characteristic imprints on the non-linear phase of collapse of halos. These imprints manifest in the non-linear power 
spectrum as well as in the halo mass function and may yield  up to 20\% deviations from $\Lambda$CDM predictions \citep{Courtin10}. In paper II (Pacaud et al, in prep.) we shall investigate how this would impact the predicted DE constraints.

\section{Evaluating the significance of $\lowercase{dn/dz}$ and $\xi$ for various survey configurations}

\label{appendixB}

We use the publicly available Pinocchio package \citep{Monaco02a,Monaco02b,Taffoni02} to generate 3D cluster 
catalogues for a given initial density field realisation and cosmology. We use
the  2.2-beta  version, that is now entirely paralleled and  available from the authors on demand.
Pinocchio, while following the 
procedure of N-body simulations, works in the Zel'dovich approximation, allowing for faster computation by several
orders of magnitude with respect to equivalent N-body simulations (in terms of mass resolution and volume probed).  
Confronting the Pinocchio realisations with the high resolution full-sky Horizon simulations  \citep{Teyssier09} (in the case of a $\Lambda$CDM model best-fit to WMAP-3 years data)  
we have checked that the Pinocchio cluster mass function 
is accurate to 10\%, and that the $2$-point correlation function can
be reliably estimated down to $10 h^{-1}{\rm Mpc}$ scale.  We observe however, a slight increase of $\xi$ around this scale, as the unresolved clusters tend to accumulate at this point.
We illustrate below our procedure considering the  C2 selection for various configurations totalling the 50 \dd\ of  {\sl Survey-{A}}.\\
 
Using Pinocchio we generate 5 cosmic volumes with different random
initial conditions for a  $\Lambda$CDM model best-fitting WMAP-5 years data
\citep{Dunkley09}. Each volume is a box of $3500 \times 3500  \times 3500$
comobile Mpc$^3$ observed from the corners, providing 8
past-lightcone octants. These octants are combined, using the
periodicity of the volumes, to finally provide 5 full-sky
past-lightcones independant from each other. The physical position
of each simulated halo is corrected for its peculiar velocity since the
correlation function is computed in redshift space. In order to
estimate the halo 2-point correlation function 5 bootstrap full-sky
lightcones are generated from the data. The angular position of each
halo is randomized 10 times to artificially create lightcones
containing 10 times more halos than the original data. The redshifts of
the original Pinocchio simulated data as well as the mass probability
distribution function are conserved in these ``random" lightcones.\\
From these lightcones we extract a large number of XXL survey realisations. We considered several survey configurations:
a single $7.07 \times 7.07$ \dd\ field, and configurations consisting of two  $5\times 5$ \dd\ and 
four $3.54\times 3.54$ \dd\ patches respectively. The last two configurations are more likely to correspond to actual observations 
since spreading patches in right ascension ensures a more efficient observation scheduling. 
Also splitting the survey into several sub-fields is usually expected to decrease the impact of the sample variance; 
an effect that we quantitatively estimate hereafter. In order to avoid large-scale correlations, edges of the 
extracted sub-fields are separated by at least 30 deg in RA and Dec. 
We also extracted survey fields covering 50x50 \dd\ for statistical comparison.
The characteristics of the different survey realisations are given in Table~\ref{simul}.\\

\begin{table}
 \centering
  \caption{Surveys extracted from the {\sl Pinocchio} simulations. Fields A0, A1, A2 pertains to different configurations of {\sl Survey-A} totalling 50 \dd. Field Z1  covers 10 000  \dd\  and is used for  statistical comparison. }
  \begin{tabular}{@{}lccc|c@{}}
  \hline\hline 
    survey configurations        &      A0      &A1 & A2 & Z1  \\
  \hline \hline
   Total surveyed area (\dd) & 50 & 50 &  50&10 000\\
   number of sub-fields & 1 & 2 & 4 &4 \\
    sub-field side (in deg) & 7.07 & 5 & 3.54  &50 \\   
    number of independent  & & & &  \\
   simulated sub-fields & 190 & 215 & 230 & 30 \\ 
 \hline \hline
\end{tabular}
\label{simul}
\end{table}

\begin{figure}
\includegraphics[width=9cm]{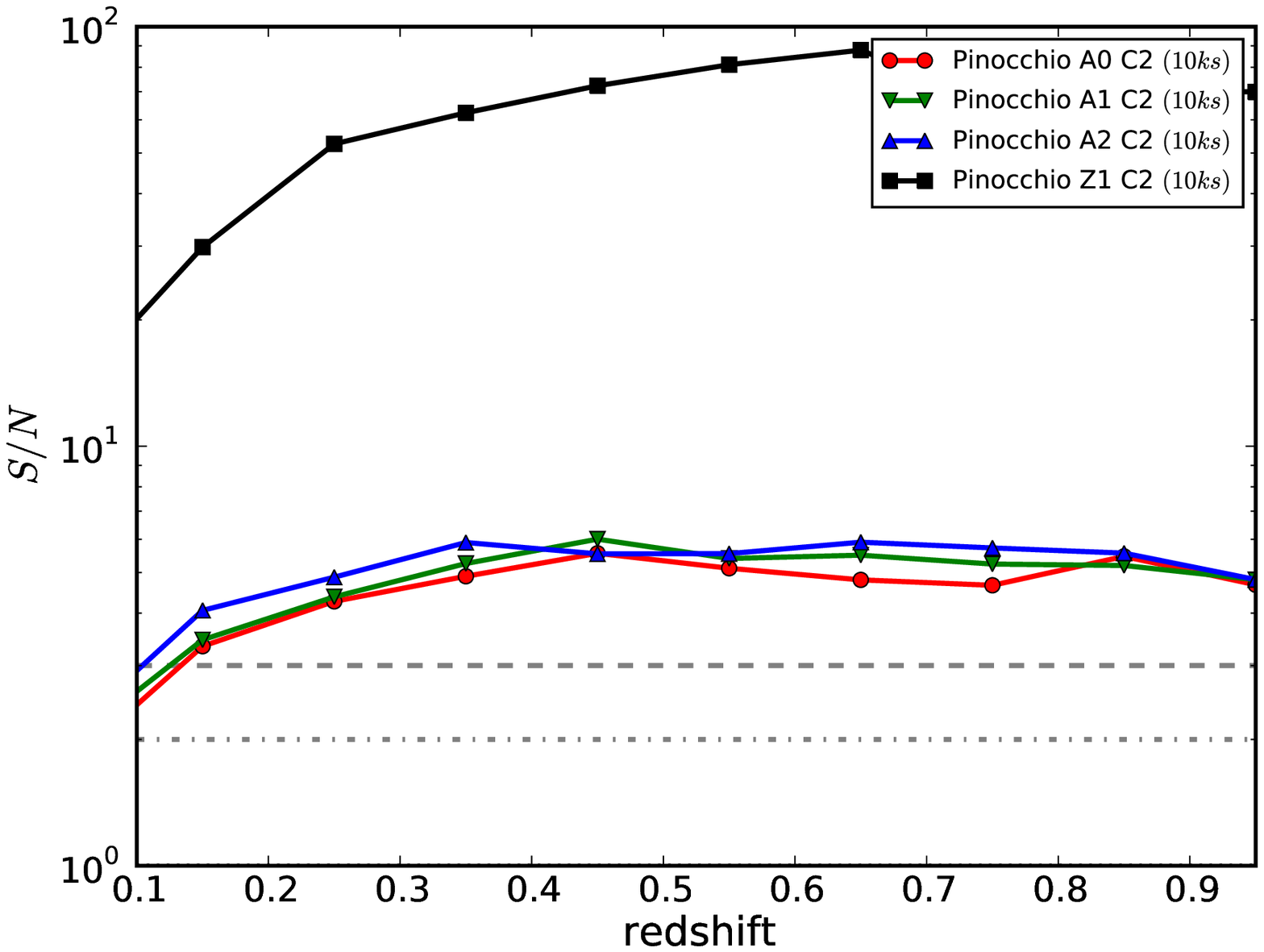}
  \caption{Signal-to-noise ratio for $dn/dz$ as a function of
  redshift, obtained for different realisations of {\sl Survey-A} and the C2 selection using
  Pinocchio simulations; bin size $\Delta{z}=0.1$ }
  \label{dndz-simulc2}
\end{figure}

\begin{figure}
\vbox{
\includegraphics[width=9cm]{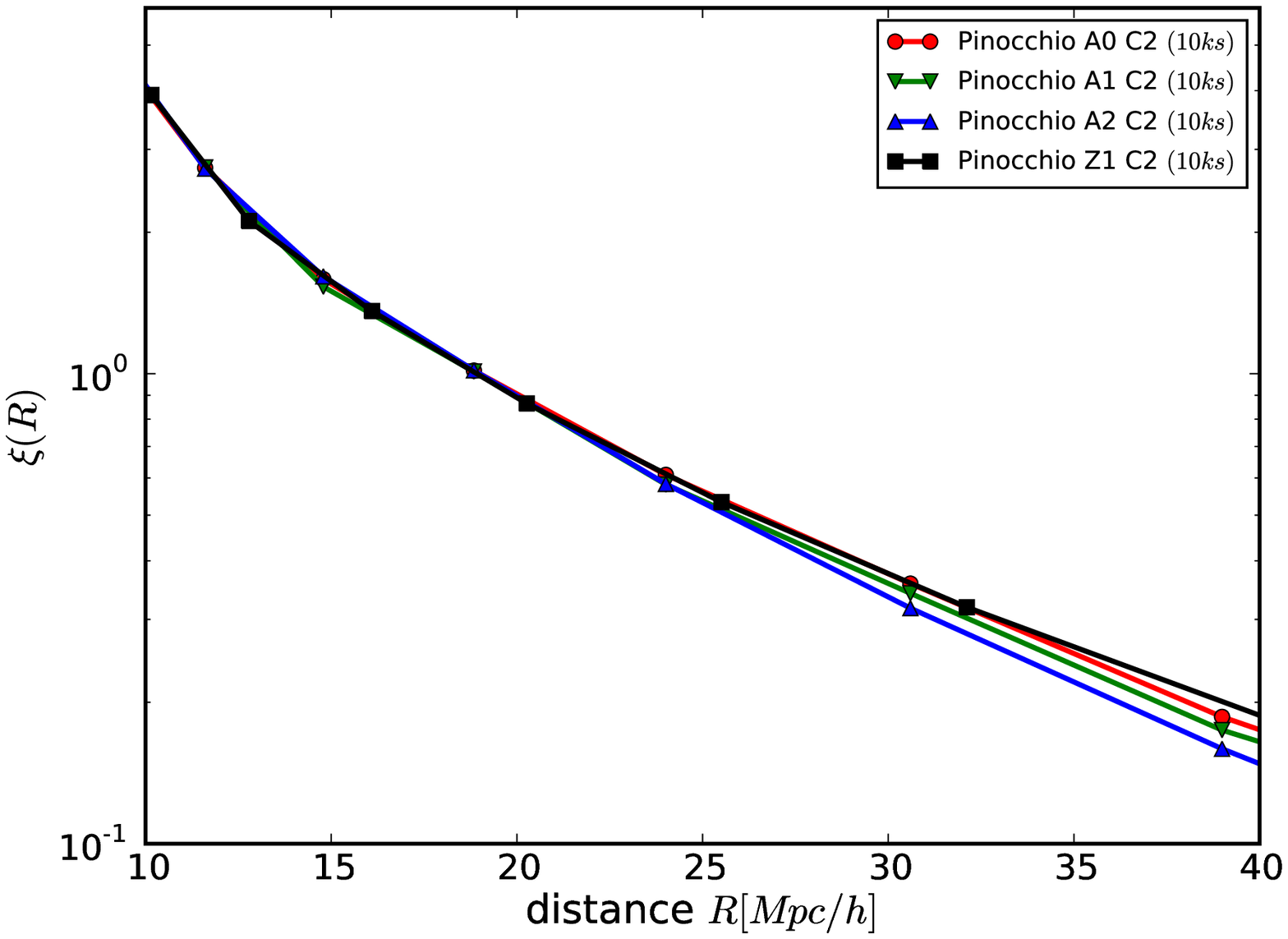}
\includegraphics[width=9cm]{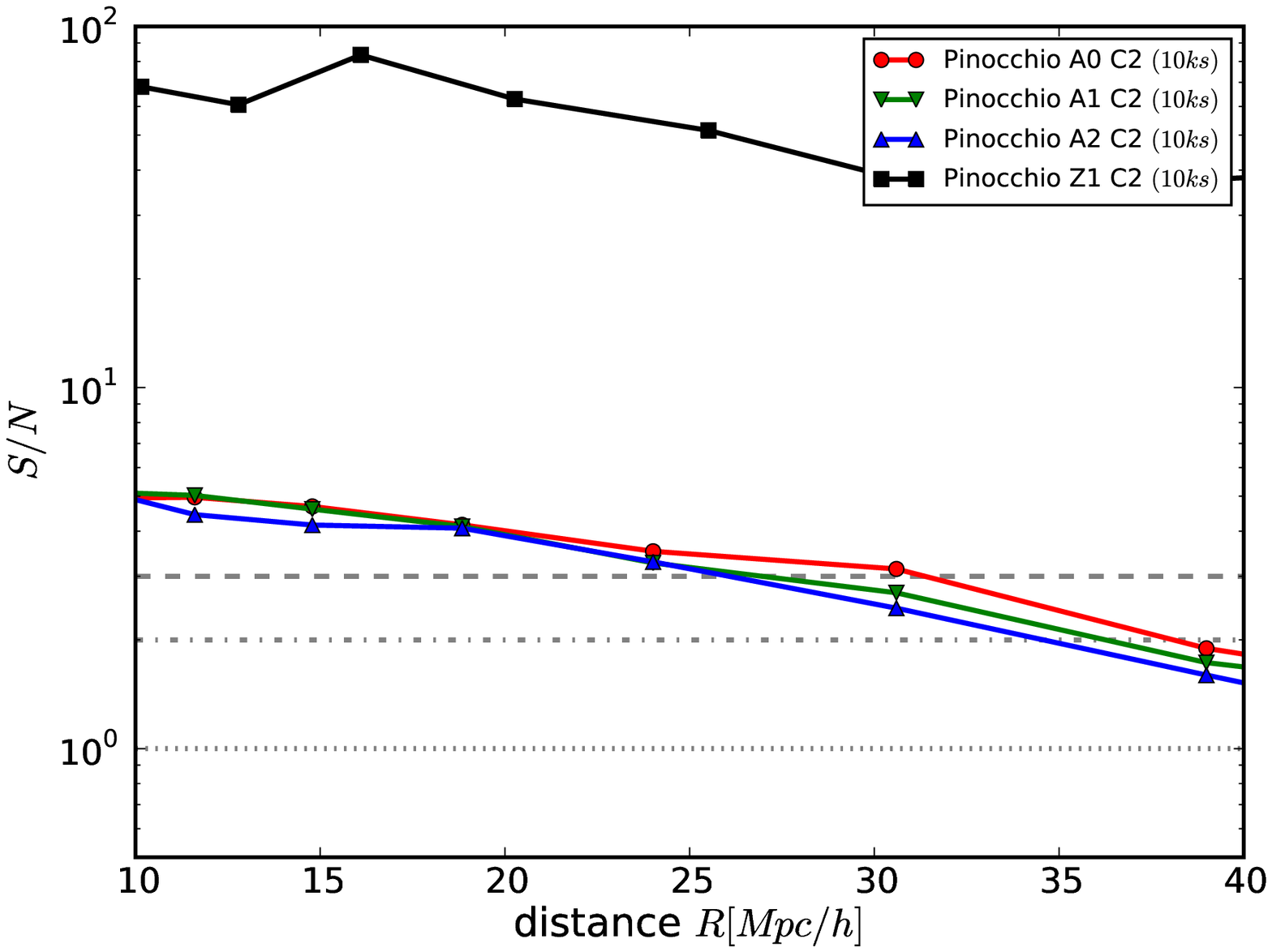}
}
  \caption{{\em Top}: 3D averaged $2$-point correlation function for
    different {\sl Survey-A} realisations extracted from the Pinocchio
    simulations using the C2 selection function. The bin
    size is $d\log R = 0.1$.  {\em Bottom}: Corresponding
    signal-to-noise ratio. }
  \label{xsi-simulc2}
\end{figure}

In each simulated sub-field, we compute $dn/dz$ and $\xi$, for a given selection. The $2$-point 
correlation fuction is measured using the estimator introduced by \cite{landy93}. The results are then 
combined according to each of the survey configurations illustrated in Table \ref{simul}; e.g. for the A2 design, 
individual $dn/dz$ are summed over the $4$ patches, while individual $\xi$ are averaged over the ensemble. 
Then for each configuration, the resulting quantities are averaged over all realisations. 
The $1\sigma$ errors about the average $dn/dz$ and $\xi$ are computed as a function of $z$ and $R$ for each of 
the 4 survey configurations, including the signal-to-noise ratio. As we describe in Section~\ref{Fisher-analysis} 
we use the estimated values of S/N to determine the experimental uncertainties necessary for the Fisher matrix analysis. 
Results are summarized in Figs.~\ref{dndz-simulc2} and
~\ref{xsi-simulc2}.  As it can be appreciated from Fig.~\ref{dndz-simulc2}, 
the cluster number counts turn out to be insensitive to the sub-field splitting 
of the survey design, i.e. a single $ 7.07 \times 7.07$  \dd\ field  (A1), two  $5 \times 5$ \dd\ sub-fields (A1) or 
four $ 3.54 \times 3.54$ \dd\ sub-fields (A2). The $2$-point correlation function appears
to be slightly dependent on the size of the sub-fields but the impact on the S/N is negligible.  
The $50 \times 50$ \dd\ Z1 reference realization  indicates that it is possible to 
reliably compute $\xi$ at least out to 40 Mpc/h for the A0, A1 or A2  configurations.
We note that 40 $h^{-1}$Mpc is slightly smaller than the comoving length encompassed by the A2 realisation at 
the survey maximum sensitivity (3.54 deg at $z=0.3$ corresponds to 53 $h^{-1}$Mpc scale).
We sample $\xi$ with a scale separation $>10 h^{-1}$Mpc because of the limited resolution 
of the Pinocchio simulations. Since cluster virial radii are on the order of $1 h^{-1}$ Mpc, this implies
that we may be loosing some power on scales of $\sim 5-10 h^{-1}$ Mpc, where mergers are expected to occur.  \\
The Pinocchio experiment  indicates the A0, A1 and A2 configurations are equivalent in terms of S/N both for $dn/dz$ and $\xi$. In the paper we consider the A2 configuration, which is for observational reason the easiest to perform, when presenting the results of the cosmological analysis.

\end{document}